\documentclass[a4paper,UKenglish,cleveref]{lipics-v2021}
\pdfoutput=1
\hideLIPIcs
\usepackage{thm-restate}
\title{Hitting Minors: Stronger Lower Bounds}
\author{Jongseo {Lee}}{School of Computing, KAIST, Daejeon, South Korea \and \url{https://leejseo.com} }{leejseo@kaist.ac.kr}{https://orcid.org/0009-0001-8936-3662}{This work was supported by the National Research Foundation
of Korea (NRF) grant funded by the Korea government (MSIT)
(RS-2025-00563533).}

\author{Roohani {Sharma}}{DIMAG, Institute for Basic Science, Daejeon, South Korea \and \url{https://sites.google.com/view/roohanisharma} }{roohani@ibs.re.kr}{https://orcid.org/0000-0003-2212-1359}{Supported by the Young Scientist Fellowship of the Institute for Basic Science (IBS-R029-Y8).}

\author{Sebastian {Wiederrecht}}{School of Computing, KAIST, Daejeon, South Korea \and \url{https://www.wiederrecht.com/} }{wiederrecht@kaist.ac.kr}{https://orcid.org/0000-0003-0462-7815}{Supported by the NRF South Korea (RS-2026-25605318) and by the Institute for Basic Science (IBS-R029-C1).}

\authorrunning{J. Lee, R. Sharma, and S. Wiederrecht}



\nolinenumbers

\usepackage{microtype}
\usepackage{tikz}
\usepackage{float}
\usetikzlibrary{fit,positioning,arrows.meta,calc,decorations.pathreplacing}
\usepackage{placeins}

\definecolor{cbblue}{RGB}{0,114,178}
\usepackage{algorithm}
\usepackage{algpseudocode}

\definecolor{Ink}{RGB}{35,39,43}
\definecolor{Muted}{RGB}{105,111,117}
\definecolor{Teal}{RGB}{0,114,178}
\definecolor{Rose}{RGB}{181,52,60}
\definecolor{Soft}{RGB}{247,248,249}
\definecolor{Mist}{RGB}{241,245,248}
\colorlet{BFConnector}{Rose}
\colorlet{BFOutput}{Teal}
\definecolor{connectorred}{RGB}{181,52,60}
\definecolor{outputblue}{RGB}{0,114,178}
\tikzset{
  vertex/.style={circle,draw=Ink,fill=Ink,inner sep=0pt,minimum size=4pt},
  output/.style={vertex,draw=Teal,fill=Teal,minimum size=4.5pt},
  hollow/.style={vertex,fill=white},
  ed/.style={draw=Ink,line width=.65pt,line cap=round},
  pathseg/.style={ed,dashed},
  BFedge/.style={line width=.65pt,line cap=round,line join=round},
  BFvertex/.style={circle,inner sep=0pt,minimum size=3.7pt},
  BFsubvertex/.style={circle,inner sep=0pt,minimum size=2.6pt},
  graph edge/.style={ed},
  compressed path/.style={dash pattern=on 3pt off 2pt},
  base vertex/.style={vertex},
  bvc vertex/.style={vertex},
  bvc survivor/.style={output},
  bvc hollow/.style={hollow,draw=Muted!65,minimum size=5pt},
  bvc edge/.style={draw=Ink!65,line width=.55pt},
  bvc ghost/.style={draw=Muted!35,dashed,line width=.55pt},
  bvc region/.style={draw=Muted!55,line width=.55pt},
  bvc bucket/.style={draw=Muted!70,dashed,line width=.55pt},
  bvc label/.style={fill=white,inner sep=1.5pt},
  bvc picture/.style={x=.84cm,y=.60cm,font=\small}
}

\newcommand{\BaseTreePicture}{%
  \begin{tikzpicture}[x=.72cm,y=.66cm,font=\small]
    \coordinate (v) at (0,0);
    \coordinate (b) at (1.9,0);
    \coordinate (a) at (5.7,0);
    \coordinate (w) at (7.6,0);
    \coordinate (z) at (1.9,-1.75);
    \coordinate (u) at (5.7,-1.75);

    \draw[graph edge] (v)--(b)--(a)--(w);
    \draw[graph edge,compressed path] (b)--(z);
    \draw[graph edge,compressed path] (a)--(u);

    \foreach \p in {v,b,a,w,z,u}
      \node[base vertex] at (\p) {};
    \foreach \p in {v,b,a,w}
      \node[above=4pt] at (\p) {$\p$};
    \node[below=4pt] at (z) {$z$};
    \node[below=4pt] at (u) {$u$};

    \node[left=7pt] at ($(b)!.5!(z)$) {$d$ edges};
    \node[right=7pt] at ($(a)!.5!(u)$) {$d$ edges};
  \end{tikzpicture}%
}

\newcommand{\ChainPaper}{%
\begin{tikzpicture}[x=.81cm,y=.46cm,baseline=(current bounding box.center),font=\small]
\path[use as bounding box] (-1.05,-4.25) rectangle (15.05,1.55);
\coordinate (v1) at (0.000,0.000);
\coordinate (b1) at (1.000,0.000);
\coordinate (a1) at (2.000,0.000);
\coordinate (w1) at (3.000,0.000);
\coordinate (z1) at (1.000,-1.550);
\coordinate (u1) at (2.000,-1.550);
\coordinate (mz1s1) at (1.000,-0.388);
\coordinate (mz1s2) at (1.000,-0.775);
\coordinate (mz1s3) at (1.000,-1.163);
\coordinate (mu1s1) at (2.000,-0.388);
\coordinate (mu1s2) at (2.000,-0.775);
\coordinate (mu1s3) at (2.000,-1.163);
\coordinate (e1) at (0.000,1.900);
\coordinate (y1) at (-0.550,0.700);
\coordinate (me1s1) at (0.000,0.475);
\coordinate (me1s2) at (0.000,0.950);
\coordinate (me1s3) at (0.000,1.425);
\coordinate (v2) at (5.700,0.000);
\coordinate (b2) at (6.700,0.000);
\coordinate (a2) at (7.700,0.000);
\coordinate (w2) at (8.700,0.000);
\coordinate (z2) at (6.700,-1.550);
\coordinate (u2) at (7.700,-1.550);
\coordinate (mz2s1) at (6.700,-0.388);
\coordinate (mz2s2) at (6.700,-0.775);
\coordinate (mz2s3) at (6.700,-1.163);
\coordinate (mu2s1) at (7.700,-0.388);
\coordinate (mu2s2) at (7.700,-0.775);
\coordinate (mu2s3) at (7.700,-1.163);
\coordinate (e2) at (5.700,1.900);
\coordinate (y2) at (5.150,0.700);
\coordinate (me2s1) at (5.700,0.475);
\coordinate (me2s2) at (5.700,0.950);
\coordinate (me2s3) at (5.700,1.425);
\coordinate (v3) at (11.400,0.000);
\coordinate (b3) at (12.400,0.000);
\coordinate (a3) at (13.400,0.000);
\coordinate (w3) at (14.400,0.000);
\coordinate (z3) at (12.400,-1.550);
\coordinate (u3) at (13.400,-1.550);
\coordinate (mz3s1) at (12.400,-0.388);
\coordinate (mz3s2) at (12.400,-0.775);
\coordinate (mz3s3) at (12.400,-1.163);
\coordinate (mu3s1) at (13.400,-0.388);
\coordinate (mu3s2) at (13.400,-0.775);
\coordinate (mu3s3) at (13.400,-1.163);
\coordinate (e3) at (11.400,1.900);
\coordinate (y3) at (10.850,0.700);
\coordinate (me3s1) at (11.400,0.475);
\coordinate (me3s2) at (11.400,0.950);
\coordinate (me3s3) at (11.400,1.425);
\coordinate (nb1) at (3.700,-2.200);
\coordinate (na1) at (4.700,-2.200);
\coordinate (nz1) at (3.700,-3.700);
\coordinate (nu1) at (4.700,-3.700);
\coordinate (nz1s1) at (3.700,-2.575);
\coordinate (nz1s2) at (3.700,-2.950);
\coordinate (nz1s3) at (3.700,-3.325);
\coordinate (nu1s1) at (4.700,-2.575);
\coordinate (nu1s2) at (4.700,-2.950);
\coordinate (nu1s3) at (4.700,-3.325);
\coordinate (nb2) at (9.400,-2.200);
\coordinate (na2) at (10.400,-2.200);
\coordinate (nz2) at (9.400,-3.700);
\coordinate (nu2) at (10.400,-3.700);
\coordinate (nz2s1) at (9.400,-2.575);
\coordinate (nz2s2) at (9.400,-2.950);
\coordinate (nz2s3) at (9.400,-3.325);
\coordinate (nu2s1) at (10.400,-2.575);
\coordinate (nu2s2) at (10.400,-2.950);
\coordinate (nu2s3) at (10.400,-3.325);
\draw[BFedge,draw=Ink] (v1) -- (b1);
\draw[BFedge,draw=Ink] (b1) -- (a1);
\draw[BFedge,draw=Ink] (a1) -- (w1);
\draw[BFedge,draw=Ink] (b1) -- (mz1s1);
\draw[BFedge,draw=Ink] (mz1s1) -- (mz1s2);
\draw[BFedge,draw=Ink] (mz1s2) -- (mz1s3);
\draw[BFedge,draw=Ink] (mz1s3) -- (z1);
\draw[BFedge,draw=Ink] (a1) -- (mu1s1);
\draw[BFedge,draw=Ink] (mu1s1) -- (mu1s2);
\draw[BFedge,draw=Ink] (mu1s2) -- (mu1s3);
\draw[BFedge,draw=Ink] (mu1s3) -- (u1);
\draw[BFedge,draw=Ink] (v2) -- (b2);
\draw[BFedge,draw=Ink] (b2) -- (a2);
\draw[BFedge,draw=Ink] (a2) -- (w2);
\draw[BFedge,draw=Ink] (b2) -- (mz2s1);
\draw[BFedge,draw=Ink] (mz2s1) -- (mz2s2);
\draw[BFedge,draw=Ink] (mz2s2) -- (mz2s3);
\draw[BFedge,draw=Ink] (mz2s3) -- (z2);
\draw[BFedge,draw=Ink] (a2) -- (mu2s1);
\draw[BFedge,draw=Ink] (mu2s1) -- (mu2s2);
\draw[BFedge,draw=Ink] (mu2s2) -- (mu2s3);
\draw[BFedge,draw=Ink] (mu2s3) -- (u2);
\draw[BFedge,draw=Ink] (v3) -- (b3);
\draw[BFedge,draw=Ink] (b3) -- (a3);
\draw[BFedge,draw=Ink] (a3) -- (w3);
\draw[BFedge,draw=Ink] (b3) -- (mz3s1);
\draw[BFedge,draw=Ink] (mz3s1) -- (mz3s2);
\draw[BFedge,draw=Ink] (mz3s2) -- (mz3s3);
\draw[BFedge,draw=Ink] (mz3s3) -- (z3);
\draw[BFedge,draw=Ink] (a3) -- (mu3s1);
\draw[BFedge,draw=Ink] (mu3s1) -- (mu3s2);
\draw[BFedge,draw=Ink] (mu3s2) -- (mu3s3);
\draw[BFedge,draw=Ink] (mu3s3) -- (u3);
\draw[BFedge,draw=BFConnector] (w1) -- (nb1);
\draw[BFedge,draw=BFConnector] (nb1) -- (na1);
\draw[BFedge,draw=BFConnector] (na1) -- (5.200,-2.800) -- (7.700,-2.800) -- (u2);
\draw[BFedge,draw=BFConnector] (nb1) -- (nz1s1);
\draw[BFedge,draw=BFConnector] (nz1s1) -- (nz1s2);
\draw[BFedge,draw=BFConnector] (nz1s2) -- (nz1s3);
\draw[BFedge,draw=BFConnector] (nz1s3) -- (nz1);
\draw[BFedge,draw=BFConnector] (na1) -- (nu1s1);
\draw[BFedge,draw=BFConnector] (nu1s1) -- (nu1s2);
\draw[BFedge,draw=BFConnector] (nu1s2) -- (nu1s3);
\draw[BFedge,draw=BFConnector] (nu1s3) -- (nu1);
\draw[BFedge,draw=BFConnector] (w2) -- (nb2);
\draw[BFedge,draw=BFConnector] (nb2) -- (na2);
\draw[BFedge,draw=BFConnector] (na2) -- (10.900,-2.800) -- (13.400,-2.800) -- (u3);
\draw[BFedge,draw=BFConnector] (nb2) -- (nz2s1);
\draw[BFedge,draw=BFConnector] (nz2s1) -- (nz2s2);
\draw[BFedge,draw=BFConnector] (nz2s2) -- (nz2s3);
\draw[BFedge,draw=BFConnector] (nz2s3) -- (nz2);
\draw[BFedge,draw=BFConnector] (na2) -- (nu2s1);
\draw[BFedge,draw=BFConnector] (nu2s1) -- (nu2s2);
\draw[BFedge,draw=BFConnector] (nu2s2) -- (nu2s3);
\draw[BFedge,draw=BFConnector] (nu2s3) -- (nu2);
\node[BFvertex,draw=Ink,fill=Ink] at (v1) {};
\node[BFvertex,draw=Ink,fill=Ink] at (b1) {};
\node[BFvertex,draw=Ink,fill=Ink] at (a1) {};
\node[BFvertex,draw=Ink,fill=Ink] at (w1) {};
\node[BFvertex,draw=Ink,fill=Ink] at (z1) {};
\node[BFvertex,draw=Ink,fill=Ink] at (u1) {};
\node[BFsubvertex,draw=Ink,fill=Ink] at (mz1s1) {};
\node[BFsubvertex,draw=Ink,fill=Ink] at (mz1s2) {};
\node[BFsubvertex,draw=Ink,fill=Ink] at (mz1s3) {};
\node[BFsubvertex,draw=Ink,fill=Ink] at (mu1s1) {};
\node[BFsubvertex,draw=Ink,fill=Ink] at (mu1s2) {};
\node[BFsubvertex,draw=Ink,fill=Ink] at (mu1s3) {};
\node[BFvertex,draw=Ink,fill=Ink] at (v2) {};
\node[BFvertex,draw=Ink,fill=Ink] at (b2) {};
\node[BFvertex,draw=Ink,fill=Ink] at (a2) {};
\node[BFvertex,draw=Ink,fill=Ink] at (w2) {};
\node[BFvertex,draw=Ink,fill=Ink] at (z2) {};
\node[BFvertex,draw=Ink,fill=Ink] at (u2) {};
\node[BFsubvertex,draw=Ink,fill=Ink] at (mz2s1) {};
\node[BFsubvertex,draw=Ink,fill=Ink] at (mz2s2) {};
\node[BFsubvertex,draw=Ink,fill=Ink] at (mz2s3) {};
\node[BFsubvertex,draw=Ink,fill=Ink] at (mu2s1) {};
\node[BFsubvertex,draw=Ink,fill=Ink] at (mu2s2) {};
\node[BFsubvertex,draw=Ink,fill=Ink] at (mu2s3) {};
\node[BFvertex,draw=Ink,fill=Ink] at (v3) {};
\node[BFvertex,draw=Ink,fill=Ink] at (b3) {};
\node[BFvertex,draw=Ink,fill=Ink] at (a3) {};
\node[BFvertex,draw=Ink,fill=Ink] at (w3) {};
\node[BFvertex,draw=Ink,fill=Ink] at (z3) {};
\node[BFvertex,draw=Ink,fill=Ink] at (u3) {};
\node[BFsubvertex,draw=Ink,fill=Ink] at (mz3s1) {};
\node[BFsubvertex,draw=Ink,fill=Ink] at (mz3s2) {};
\node[BFsubvertex,draw=Ink,fill=Ink] at (mz3s3) {};
\node[BFsubvertex,draw=Ink,fill=Ink] at (mu3s1) {};
\node[BFsubvertex,draw=Ink,fill=Ink] at (mu3s2) {};
\node[BFsubvertex,draw=Ink,fill=Ink] at (mu3s3) {};
\node[BFvertex,draw=BFConnector,fill=BFConnector] at (nb1) {};
\node[BFvertex,draw=BFConnector,fill=BFConnector] at (na1) {};
\node[BFvertex,draw=BFConnector,fill=BFConnector] at (nz1) {};
\node[BFvertex,draw=BFConnector,fill=BFConnector] at (nu1) {};
\node[BFsubvertex,draw=BFConnector,fill=BFConnector] at (nz1s1) {};
\node[BFsubvertex,draw=BFConnector,fill=BFConnector] at (nz1s2) {};
\node[BFsubvertex,draw=BFConnector,fill=BFConnector] at (nz1s3) {};
\node[BFsubvertex,draw=BFConnector,fill=BFConnector] at (nu1s1) {};
\node[BFsubvertex,draw=BFConnector,fill=BFConnector] at (nu1s2) {};
\node[BFsubvertex,draw=BFConnector,fill=BFConnector] at (nu1s3) {};
\node[BFvertex,draw=BFConnector,fill=BFConnector] at (nb2) {};
\node[BFvertex,draw=BFConnector,fill=BFConnector] at (na2) {};
\node[BFvertex,draw=BFConnector,fill=BFConnector] at (nz2) {};
\node[BFvertex,draw=BFConnector,fill=BFConnector] at (nu2) {};
\node[BFsubvertex,draw=BFConnector,fill=BFConnector] at (nz2s1) {};
\node[BFsubvertex,draw=BFConnector,fill=BFConnector] at (nz2s2) {};
\node[BFsubvertex,draw=BFConnector,fill=BFConnector] at (nz2s3) {};
\node[BFsubvertex,draw=BFConnector,fill=BFConnector] at (nu2s1) {};
\node[BFsubvertex,draw=BFConnector,fill=BFConnector] at (nu2s2) {};
\node[BFsubvertex,draw=BFConnector,fill=BFConnector] at (nu2s3) {};
\node[below left=3pt,text=Ink,inner sep=1pt] at (v1) {$v_1$};
\node[above=3pt,text=Ink,inner sep=1pt] at (b1) {$b_1$};
\node[above=3pt,text=Ink,inner sep=1pt] at (a1) {$a_1$};
\node[above=3pt,text=Ink,inner sep=1pt] at (w1) {$w_1$};
\node[below=3pt,text=Ink,inner sep=1pt] at (z1) {$z_1$};
\node[below=3pt,text=Ink,inner sep=1pt] at (u1) {$u_1$};
\node[below left=3pt,text=Ink,inner sep=1pt] at (v2) {$v_2$};
\node[above=3pt,text=Ink,inner sep=1pt] at (b2) {$b_2$};
\node[above=3pt,text=Ink,inner sep=1pt] at (a2) {$a_2$};
\node[above=3pt,text=Ink,inner sep=1pt] at (w2) {$w_2$};
\node[below=3pt,text=Ink,inner sep=1pt] at (z2) {$z_2$};
\node[right=5pt,text=Ink,inner sep=1pt] at (u2) {$u_2$};
\node[below left=3pt,text=Ink,inner sep=1pt] at (v3) {$v_3$};
\node[above=3pt,text=Ink,inner sep=1pt] at (b3) {$b_3$};
\node[above=3pt,text=Ink,inner sep=1pt] at (a3) {$a_3$};
\node[above=3pt,text=Ink,inner sep=1pt] at (w3) {$w_3$};
\node[below=3pt,text=Ink,inner sep=1pt] at (z3) {$z_3$};
\node[right=5pt,text=Ink,inner sep=1pt] at (u3) {$u_3$};
\node[above left=3pt,text=BFConnector,inner sep=1pt] at (nb1) {$b^{\prime}_1$};
\node[above right=3pt,text=BFConnector,inner sep=1pt] at (na1) {$a^{\prime}_1$};
\node[below=3pt,text=BFConnector,inner sep=1pt] at (nz1) {$z^{\prime}_1$};
\node[below=3pt,text=BFConnector,inner sep=1pt] at (nu1) {$u^{\prime}_1$};
\node[above left=3pt,text=BFConnector,inner sep=1pt] at (nb2) {$b^{\prime}_2$};
\node[above right=3pt,text=BFConnector,inner sep=1pt] at (na2) {$a^{\prime}_2$};
\node[below=3pt,text=BFConnector,inner sep=1pt] at (nz2) {$z^{\prime}_2$};
\node[below=3pt,text=BFConnector,inner sep=1pt] at (nu2) {$u^{\prime}_2$};
\node[text=Muted] at (1.600,1.35) {$M_1$};
\node[text=Muted] at (7.300,1.35) {$M_2$};
\node[text=Muted] at (13.000,1.35) {$M_3$};
\node[text=BFConnector] at (4.200,-1.53) {$N_1$};
\node[text=BFConnector] at (9.900,-1.53) {$N_2$};
\end{tikzpicture}
}

\newcommand{\SelectorPaper}{%
\begin{tikzpicture}[x=.81cm,y=.46cm,baseline=(current bounding box.center),font=\small]
\path[use as bounding box] (-1.05,-4.25) rectangle (15.05,2.5);
\coordinate (v1) at (0.000,0.000);
\coordinate (b1) at (1.000,0.000);
\coordinate (a1) at (2.000,0.000);
\coordinate (w1) at (3.000,0.000);
\coordinate (z1) at (1.000,-1.550);
\coordinate (u1) at (2.000,-1.550);
\coordinate (mz1s1) at (1.000,-0.388);
\coordinate (mz1s2) at (1.000,-0.775);
\coordinate (mz1s3) at (1.000,-1.163);
\coordinate (mu1s1) at (2.000,-0.388);
\coordinate (mu1s2) at (2.000,-0.775);
\coordinate (mu1s3) at (2.000,-1.163);
\coordinate (e1) at (0.000,1.900);
\coordinate (y1) at (-0.550,0.700);
\coordinate (me1s1) at (0.000,0.475);
\coordinate (me1s2) at (0.000,0.950);
\coordinate (me1s3) at (0.000,1.425);
\coordinate (v2) at (5.700,0.000);
\coordinate (b2) at (6.700,0.000);
\coordinate (a2) at (7.700,0.000);
\coordinate (w2) at (8.700,0.000);
\coordinate (z2) at (6.700,-1.550);
\coordinate (u2) at (7.700,-1.550);
\coordinate (mz2s1) at (6.700,-0.388);
\coordinate (mz2s2) at (6.700,-0.775);
\coordinate (mz2s3) at (6.700,-1.163);
\coordinate (mu2s1) at (7.700,-0.388);
\coordinate (mu2s2) at (7.700,-0.775);
\coordinate (mu2s3) at (7.700,-1.163);
\coordinate (e2) at (5.700,1.900);
\coordinate (y2) at (5.150,0.700);
\coordinate (me2s1) at (5.700,0.475);
\coordinate (me2s2) at (5.700,0.950);
\coordinate (me2s3) at (5.700,1.425);
\coordinate (v3) at (11.400,0.000);
\coordinate (b3) at (12.400,0.000);
\coordinate (a3) at (13.400,0.000);
\coordinate (w3) at (14.400,0.000);
\coordinate (z3) at (12.400,-1.550);
\coordinate (u3) at (13.400,-1.550);
\coordinate (mz3s1) at (12.400,-0.388);
\coordinate (mz3s2) at (12.400,-0.775);
\coordinate (mz3s3) at (12.400,-1.163);
\coordinate (mu3s1) at (13.400,-0.388);
\coordinate (mu3s2) at (13.400,-0.775);
\coordinate (mu3s3) at (13.400,-1.163);
\coordinate (e3) at (11.400,1.900);
\coordinate (y3) at (10.850,0.700);
\coordinate (me3s1) at (11.400,0.475);
\coordinate (me3s2) at (11.400,0.950);
\coordinate (me3s3) at (11.400,1.425);
\coordinate (nb1) at (3.700,-2.200);
\coordinate (na1) at (4.700,-2.200);
\coordinate (nz1) at (3.700,-3.700);
\coordinate (nu1) at (4.700,-3.700);
\coordinate (nz1s1) at (3.700,-2.575);
\coordinate (nz1s2) at (3.700,-2.950);
\coordinate (nz1s3) at (3.700,-3.325);
\coordinate (nu1s1) at (4.700,-2.575);
\coordinate (nu1s2) at (4.700,-2.950);
\coordinate (nu1s3) at (4.700,-3.325);
\coordinate (nb2) at (9.400,-2.200);
\coordinate (na2) at (10.400,-2.200);
\coordinate (nz2) at (9.400,-3.700);
\coordinate (nu2) at (10.400,-3.700);
\coordinate (nz2s1) at (9.400,-2.575);
\coordinate (nz2s2) at (9.400,-2.950);
\coordinate (nz2s3) at (9.400,-3.325);
\coordinate (nu2s1) at (10.400,-2.575);
\coordinate (nu2s2) at (10.400,-2.950);
\coordinate (nu2s3) at (10.400,-3.325);
\draw[BFedge,draw=Ink] (v1) -- (b1);
\draw[BFedge,draw=Ink] (b1) -- (a1);
\draw[BFedge,draw=Ink] (a1) -- (w1);
\draw[BFedge,draw=Ink] (b1) -- (mz1s1);
\draw[BFedge,draw=Ink] (mz1s1) -- (mz1s2);
\draw[BFedge,draw=Ink] (mz1s2) -- (mz1s3);
\draw[BFedge,draw=Ink] (mz1s3) -- (z1);
\draw[BFedge,draw=Ink] (a1) -- (mu1s1);
\draw[BFedge,draw=Ink] (mu1s1) -- (mu1s2);
\draw[BFedge,draw=Ink] (mu1s2) -- (mu1s3);
\draw[BFedge,draw=Ink] (mu1s3) -- (u1);
\draw[BFedge,draw=BFOutput] (v1) -- (me1s1);
\draw[BFedge,draw=BFOutput] (me1s1) -- (me1s2);
\draw[BFedge,draw=BFOutput] (me1s2) -- (me1s3);
\draw[BFedge,draw=BFOutput] (me1s3) -- (e1);
\draw[BFedge,draw=BFOutput] (v1) -- (y1);
\draw[BFedge,draw=Ink] (v2) -- (b2);
\draw[BFedge,draw=Ink] (b2) -- (a2);
\draw[BFedge,draw=Ink] (a2) -- (w2);
\draw[BFedge,draw=Ink] (b2) -- (mz2s1);
\draw[BFedge,draw=Ink] (mz2s1) -- (mz2s2);
\draw[BFedge,draw=Ink] (mz2s2) -- (mz2s3);
\draw[BFedge,draw=Ink] (mz2s3) -- (z2);
\draw[BFedge,draw=Ink] (a2) -- (mu2s1);
\draw[BFedge,draw=Ink] (mu2s1) -- (mu2s2);
\draw[BFedge,draw=Ink] (mu2s2) -- (mu2s3);
\draw[BFedge,draw=Ink] (mu2s3) -- (u2);
\draw[BFedge,draw=BFOutput] (v2) -- (me2s1);
\draw[BFedge,draw=BFOutput] (me2s1) -- (me2s2);
\draw[BFedge,draw=BFOutput] (me2s2) -- (me2s3);
\draw[BFedge,draw=BFOutput] (me2s3) -- (e2);
\draw[BFedge,draw=BFOutput] (v2) -- (y2);
\draw[BFedge,draw=Ink] (v3) -- (b3);
\draw[BFedge,draw=Ink] (b3) -- (a3);
\draw[BFedge,draw=Ink] (a3) -- (w3);
\draw[BFedge,draw=Ink] (b3) -- (mz3s1);
\draw[BFedge,draw=Ink] (mz3s1) -- (mz3s2);
\draw[BFedge,draw=Ink] (mz3s2) -- (mz3s3);
\draw[BFedge,draw=Ink] (mz3s3) -- (z3);
\draw[BFedge,draw=Ink] (a3) -- (mu3s1);
\draw[BFedge,draw=Ink] (mu3s1) -- (mu3s2);
\draw[BFedge,draw=Ink] (mu3s2) -- (mu3s3);
\draw[BFedge,draw=Ink] (mu3s3) -- (u3);
\draw[BFedge,draw=BFOutput] (v3) -- (me3s1);
\draw[BFedge,draw=BFOutput] (me3s1) -- (me3s2);
\draw[BFedge,draw=BFOutput] (me3s2) -- (me3s3);
\draw[BFedge,draw=BFOutput] (me3s3) -- (e3);
\draw[BFedge,draw=BFOutput] (v3) -- (y3);
\draw[BFedge,draw=BFConnector] (w1) -- (nb1);
\draw[BFedge,draw=BFConnector] (nb1) -- (na1);
\draw[BFedge,draw=BFConnector] (na1) -- (5.200,-2.800) -- (7.700,-2.800) -- (u2);
\draw[BFedge,draw=BFConnector] (nb1) -- (nz1s1);
\draw[BFedge,draw=BFConnector] (nz1s1) -- (nz1s2);
\draw[BFedge,draw=BFConnector] (nz1s2) -- (nz1s3);
\draw[BFedge,draw=BFConnector] (nz1s3) -- (nz1);
\draw[BFedge,draw=BFConnector] (na1) -- (nu1s1);
\draw[BFedge,draw=BFConnector] (nu1s1) -- (nu1s2);
\draw[BFedge,draw=BFConnector] (nu1s2) -- (nu1s3);
\draw[BFedge,draw=BFConnector] (nu1s3) -- (nu1);
\draw[BFedge,draw=BFConnector] (w2) -- (nb2);
\draw[BFedge,draw=BFConnector] (nb2) -- (na2);
\draw[BFedge,draw=BFConnector] (na2) -- (10.900,-2.800) -- (13.400,-2.800) -- (u3);
\draw[BFedge,draw=BFConnector] (nb2) -- (nz2s1);
\draw[BFedge,draw=BFConnector] (nz2s1) -- (nz2s2);
\draw[BFedge,draw=BFConnector] (nz2s2) -- (nz2s3);
\draw[BFedge,draw=BFConnector] (nz2s3) -- (nz2);
\draw[BFedge,draw=BFConnector] (na2) -- (nu2s1);
\draw[BFedge,draw=BFConnector] (nu2s1) -- (nu2s2);
\draw[BFedge,draw=BFConnector] (nu2s2) -- (nu2s3);
\draw[BFedge,draw=BFConnector] (nu2s3) -- (nu2);
\node[BFvertex,draw=Ink,fill=Ink] at (v1) {};
\node[BFvertex,draw=Ink,fill=Ink] at (b1) {};
\node[BFvertex,draw=Ink,fill=Ink] at (a1) {};
\node[BFvertex,draw=Ink,fill=Ink] at (w1) {};
\node[BFvertex,draw=Ink,fill=Ink] at (z1) {};
\node[BFvertex,draw=Ink,fill=Ink] at (u1) {};
\node[BFsubvertex,draw=Ink,fill=Ink] at (mz1s1) {};
\node[BFsubvertex,draw=Ink,fill=Ink] at (mz1s2) {};
\node[BFsubvertex,draw=Ink,fill=Ink] at (mz1s3) {};
\node[BFsubvertex,draw=Ink,fill=Ink] at (mu1s1) {};
\node[BFsubvertex,draw=Ink,fill=Ink] at (mu1s2) {};
\node[BFsubvertex,draw=Ink,fill=Ink] at (mu1s3) {};
\node[BFvertex,draw=BFOutput,fill=BFOutput] at (e1) {};
\node[BFvertex,draw=BFOutput,fill=BFOutput] at (y1) {};
\node[BFsubvertex,draw=BFOutput,fill=BFOutput] at (me1s1) {};
\node[BFsubvertex,draw=BFOutput,fill=BFOutput] at (me1s2) {};
\node[BFsubvertex,draw=BFOutput,fill=BFOutput] at (me1s3) {};
\node[BFvertex,draw=Ink,fill=Ink] at (v2) {};
\node[BFvertex,draw=Ink,fill=Ink] at (b2) {};
\node[BFvertex,draw=Ink,fill=Ink] at (a2) {};
\node[BFvertex,draw=Ink,fill=Ink] at (w2) {};
\node[BFvertex,draw=Ink,fill=Ink] at (z2) {};
\node[BFvertex,draw=Ink,fill=Ink] at (u2) {};
\node[BFsubvertex,draw=Ink,fill=Ink] at (mz2s1) {};
\node[BFsubvertex,draw=Ink,fill=Ink] at (mz2s2) {};
\node[BFsubvertex,draw=Ink,fill=Ink] at (mz2s3) {};
\node[BFsubvertex,draw=Ink,fill=Ink] at (mu2s1) {};
\node[BFsubvertex,draw=Ink,fill=Ink] at (mu2s2) {};
\node[BFsubvertex,draw=Ink,fill=Ink] at (mu2s3) {};
\node[BFvertex,draw=BFOutput,fill=BFOutput] at (e2) {};
\node[BFvertex,draw=BFOutput,fill=BFOutput] at (y2) {};
\node[BFsubvertex,draw=BFOutput,fill=BFOutput] at (me2s1) {};
\node[BFsubvertex,draw=BFOutput,fill=BFOutput] at (me2s2) {};
\node[BFsubvertex,draw=BFOutput,fill=BFOutput] at (me2s3) {};
\node[BFvertex,draw=Ink,fill=Ink] at (v3) {};
\node[BFvertex,draw=Ink,fill=Ink] at (b3) {};
\node[BFvertex,draw=Ink,fill=Ink] at (a3) {};
\node[BFvertex,draw=Ink,fill=Ink] at (w3) {};
\node[BFvertex,draw=Ink,fill=Ink] at (z3) {};
\node[BFvertex,draw=Ink,fill=Ink] at (u3) {};
\node[BFsubvertex,draw=Ink,fill=Ink] at (mz3s1) {};
\node[BFsubvertex,draw=Ink,fill=Ink] at (mz3s2) {};
\node[BFsubvertex,draw=Ink,fill=Ink] at (mz3s3) {};
\node[BFsubvertex,draw=Ink,fill=Ink] at (mu3s1) {};
\node[BFsubvertex,draw=Ink,fill=Ink] at (mu3s2) {};
\node[BFsubvertex,draw=Ink,fill=Ink] at (mu3s3) {};
\node[BFvertex,draw=BFOutput,fill=BFOutput] at (e3) {};
\node[BFvertex,draw=BFOutput,fill=BFOutput] at (y3) {};
\node[BFsubvertex,draw=BFOutput,fill=BFOutput] at (me3s1) {};
\node[BFsubvertex,draw=BFOutput,fill=BFOutput] at (me3s2) {};
\node[BFsubvertex,draw=BFOutput,fill=BFOutput] at (me3s3) {};
\node[BFvertex,draw=BFConnector,fill=BFConnector] at (nb1) {};
\node[BFvertex,draw=BFConnector,fill=BFConnector] at (na1) {};
\node[BFvertex,draw=BFConnector,fill=BFConnector] at (nz1) {};
\node[BFvertex,draw=BFConnector,fill=BFConnector] at (nu1) {};
\node[BFsubvertex,draw=BFConnector,fill=BFConnector] at (nz1s1) {};
\node[BFsubvertex,draw=BFConnector,fill=BFConnector] at (nz1s2) {};
\node[BFsubvertex,draw=BFConnector,fill=BFConnector] at (nz1s3) {};
\node[BFsubvertex,draw=BFConnector,fill=BFConnector] at (nu1s1) {};
\node[BFsubvertex,draw=BFConnector,fill=BFConnector] at (nu1s2) {};
\node[BFsubvertex,draw=BFConnector,fill=BFConnector] at (nu1s3) {};
\node[BFvertex,draw=BFConnector,fill=BFConnector] at (nb2) {};
\node[BFvertex,draw=BFConnector,fill=BFConnector] at (na2) {};
\node[BFvertex,draw=BFConnector,fill=BFConnector] at (nz2) {};
\node[BFvertex,draw=BFConnector,fill=BFConnector] at (nu2) {};
\node[BFsubvertex,draw=BFConnector,fill=BFConnector] at (nz2s1) {};
\node[BFsubvertex,draw=BFConnector,fill=BFConnector] at (nz2s2) {};
\node[BFsubvertex,draw=BFConnector,fill=BFConnector] at (nz2s3) {};
\node[BFsubvertex,draw=BFConnector,fill=BFConnector] at (nu2s1) {};
\node[BFsubvertex,draw=BFConnector,fill=BFConnector] at (nu2s2) {};
\node[BFsubvertex,draw=BFConnector,fill=BFConnector] at (nu2s3) {};
\node[below left=3pt,text=Ink,inner sep=1pt] at (v1) {$v_1$};
\node[above=3pt,text=Ink,inner sep=1pt] at (b1) {$b_1$};
\node[above=3pt,text=Ink,inner sep=1pt] at (a1) {$a_1$};
\node[above=3pt,text=Ink,inner sep=1pt] at (w1) {$w_1$};
\node[below=3pt,text=Ink,inner sep=1pt] at (z1) {$z_1$};
\node[below=3pt,text=Ink,inner sep=1pt] at (u1) {$u_1$};
\node[above=3pt,text=BFOutput,inner sep=1pt] at (e1) {$e_1$};
\node[left=3pt,text=BFOutput,inner sep=1pt] at (y1) {$y_1$};
\node[below left=3pt,text=Ink,inner sep=1pt] at (v2) {$v_2$};
\node[above=3pt,text=Ink,inner sep=1pt] at (b2) {$b_2$};
\node[above=3pt,text=Ink,inner sep=1pt] at (a2) {$a_2$};
\node[above=3pt,text=Ink,inner sep=1pt] at (w2) {$w_2$};
\node[below=3pt,text=Ink,inner sep=1pt] at (z2) {$z_2$};
\node[right=5pt,text=Ink,inner sep=1pt] at (u2) {$u_2$};
\node[above=3pt,text=BFOutput,inner sep=1pt] at (e2) {$e_2$};
\node[left=3pt,text=BFOutput,inner sep=1pt] at (y2) {$y_2$};
\node[below left=3pt,text=Ink,inner sep=1pt] at (v3) {$v_3$};
\node[above=3pt,text=Ink,inner sep=1pt] at (b3) {$b_3$};
\node[above=3pt,text=Ink,inner sep=1pt] at (a3) {$a_3$};
\node[above=3pt,text=Ink,inner sep=1pt] at (w3) {$w_3$};
\node[below=3pt,text=Ink,inner sep=1pt] at (z3) {$z_3$};
\node[right=5pt,text=Ink,inner sep=1pt] at (u3) {$u_3$};
\node[above=3pt,text=BFOutput,inner sep=1pt] at (e3) {$e_3$};
\node[left=3pt,text=BFOutput,inner sep=1pt] at (y3) {$y_3$};
\node[above left=3pt,text=BFConnector,inner sep=1pt] at (nb1) {$b^{\prime}_1$};
\node[above right=3pt,text=BFConnector,inner sep=1pt] at (na1) {$a^{\prime}_1$};
\node[below=3pt,text=BFConnector,inner sep=1pt] at (nz1) {$z^{\prime}_1$};
\node[below=3pt,text=BFConnector,inner sep=1pt] at (nu1) {$u^{\prime}_1$};
\node[above left=3pt,text=BFConnector,inner sep=1pt] at (nb2) {$b^{\prime}_2$};
\node[above right=3pt,text=BFConnector,inner sep=1pt] at (na2) {$a^{\prime}_2$};
\node[below=3pt,text=BFConnector,inner sep=1pt] at (nz2) {$z^{\prime}_2$};
\node[below=3pt,text=BFConnector,inner sep=1pt] at (nu2) {$u^{\prime}_2$};
\node[text=Muted] at (1.600,1.35) {$M_1$};
\node[text=Muted] at (7.300,1.35) {$M_2$};
\node[text=Muted] at (13.000,1.35) {$M_3$};
\node[text=BFConnector] at (4.200,-1.53) {$N_1$};
\node[text=BFConnector] at (9.900,-1.53) {$N_2$};
\end{tikzpicture}
}

\newcommand{\CanonicalPaper}{%
\begin{tikzpicture}[x=.81cm,y=.46cm,baseline=(current bounding box.center),font=\small]
\path[use as bounding box] (-1.05,-4.25) rectangle (15.05,2.5);
\coordinate (v1) at (0.000,0.000);
\coordinate (b1) at (1.000,0.000);
\coordinate (a1) at (2.000,0.000);
\coordinate (w1) at (3.000,0.000);
\coordinate (z1) at (1.000,-1.550);
\coordinate (u1) at (2.000,-1.550);
\coordinate (mz1s1) at (1.000,-0.388);
\coordinate (mz1s2) at (1.000,-0.775);
\coordinate (mz1s3) at (1.000,-1.163);
\coordinate (mu1s1) at (2.000,-0.388);
\coordinate (mu1s2) at (2.000,-0.775);
\coordinate (mu1s3) at (2.000,-1.163);
\coordinate (e1) at (0.000,1.900);
\coordinate (y1) at (-0.550,0.700);
\coordinate (me1s1) at (0.000,0.475);
\coordinate (me1s2) at (0.000,0.950);
\coordinate (me1s3) at (0.000,1.425);
\coordinate (v2) at (5.700,0.000);
\coordinate (b2) at (6.700,0.000);
\coordinate (a2) at (7.700,0.000);
\coordinate (w2) at (8.700,0.000);
\coordinate (z2) at (6.700,-1.550);
\coordinate (u2) at (7.700,-1.550);
\coordinate (mz2s1) at (6.700,-0.388);
\coordinate (mz2s2) at (6.700,-0.775);
\coordinate (mz2s3) at (6.700,-1.163);
\coordinate (mu2s1) at (7.700,-0.388);
\coordinate (mu2s2) at (7.700,-0.775);
\coordinate (mu2s3) at (7.700,-1.163);
\coordinate (e2) at (5.700,1.900);
\coordinate (y2) at (5.150,0.700);
\coordinate (me2s1) at (5.700,0.475);
\coordinate (me2s2) at (5.700,0.950);
\coordinate (me2s3) at (5.700,1.425);
\coordinate (v3) at (11.400,0.000);
\coordinate (b3) at (12.400,0.000);
\coordinate (a3) at (13.400,0.000);
\coordinate (w3) at (14.400,0.000);
\coordinate (z3) at (12.400,-1.550);
\coordinate (u3) at (13.400,-1.550);
\coordinate (mz3s1) at (12.400,-0.388);
\coordinate (mz3s2) at (12.400,-0.775);
\coordinate (mz3s3) at (12.400,-1.163);
\coordinate (mu3s1) at (13.400,-0.388);
\coordinate (mu3s2) at (13.400,-0.775);
\coordinate (mu3s3) at (13.400,-1.163);
\coordinate (e3) at (11.400,1.900);
\coordinate (y3) at (10.850,0.700);
\coordinate (me3s1) at (11.400,0.475);
\coordinate (me3s2) at (11.400,0.950);
\coordinate (me3s3) at (11.400,1.425);
\coordinate (nb1) at (3.700,-2.200);
\coordinate (na1) at (4.700,-2.200);
\coordinate (nz1) at (3.700,-3.700);
\coordinate (nu1) at (4.700,-3.700);
\coordinate (nz1s1) at (3.700,-2.575);
\coordinate (nz1s2) at (3.700,-2.950);
\coordinate (nz1s3) at (3.700,-3.325);
\coordinate (nu1s1) at (4.700,-2.575);
\coordinate (nu1s2) at (4.700,-2.950);
\coordinate (nu1s3) at (4.700,-3.325);
\coordinate (nb2) at (9.400,-2.200);
\coordinate (na2) at (10.400,-2.200);
\coordinate (nz2) at (9.400,-3.700);
\coordinate (nu2) at (10.400,-3.700);
\coordinate (nz2s1) at (9.400,-2.575);
\coordinate (nz2s2) at (9.400,-2.950);
\coordinate (nz2s3) at (9.400,-3.325);
\coordinate (nu2s1) at (10.400,-2.575);
\coordinate (nu2s2) at (10.400,-2.950);
\coordinate (nu2s3) at (10.400,-3.325);
\draw[BFedge,draw=Ink,opacity=.28] (v1) -- (b1);
\draw[BFedge,draw=Ink,opacity=.28] (b1) -- (a1);
\draw[BFedge,draw=Ink,opacity=.28] (a1) -- (w1);
\draw[BFedge,draw=Ink,opacity=.28] (b1) -- (mz1s1);
\draw[BFedge,draw=Ink] (mz1s1) -- (mz1s2);
\draw[BFedge,draw=Ink] (mz1s2) -- (mz1s3);
\draw[BFedge,draw=Ink] (mz1s3) -- (z1);
\draw[BFedge,draw=Ink] (a1) -- (mu1s1);
\draw[BFedge,draw=Ink] (mu1s1) -- (mu1s2);
\draw[BFedge,draw=Ink] (mu1s2) -- (mu1s3);
\draw[BFedge,draw=Ink] (mu1s3) -- (u1);
\draw[BFedge,draw=BFOutput] (v1) -- (me1s1);
\draw[BFedge,draw=BFOutput] (me1s1) -- (me1s2);
\draw[BFedge,draw=BFOutput] (me1s2) -- (me1s3);
\draw[BFedge,draw=BFOutput] (me1s3) -- (e1);
\draw[BFedge,draw=BFOutput,opacity=.28] (v1) -- (y1);
\draw[BFedge,draw=Ink,opacity=.28] (v2) -- (b2);
\draw[BFedge,draw=Ink,opacity=.28] (b2) -- (a2);
\draw[BFedge,draw=Ink,opacity=.28] (a2) -- (w2);
\draw[BFedge,draw=Ink] (b2) -- (mz2s1);
\draw[BFedge,draw=Ink] (mz2s1) -- (mz2s2);
\draw[BFedge,draw=Ink] (mz2s2) -- (mz2s3);
\draw[BFedge,draw=Ink] (mz2s3) -- (z2);
\draw[BFedge,draw=Ink,opacity=.28] (a2) -- (mu2s1);
\draw[BFedge,draw=Ink] (mu2s1) -- (mu2s2);
\draw[BFedge,draw=Ink] (mu2s2) -- (mu2s3);
\draw[BFedge,draw=Ink] (mu2s3) -- (u2);
\draw[BFedge,draw=BFOutput,opacity=.28] (v2) -- (me2s1);
\draw[BFedge,draw=BFOutput] (me2s1) -- (me2s2);
\draw[BFedge,draw=BFOutput] (me2s2) -- (me2s3);
\draw[BFedge,draw=BFOutput] (me2s3) -- (e2);
\draw[BFedge,draw=BFOutput,opacity=.28] (v2) -- (y2);
\draw[BFedge,draw=Ink,opacity=.28] (v3) -- (b3);
\draw[BFedge,draw=Ink,opacity=.28] (b3) -- (a3);
\draw[BFedge,draw=Ink] (a3) -- (w3);
\draw[BFedge,draw=Ink,opacity=.28] (b3) -- (mz3s1);
\draw[BFedge,draw=Ink] (mz3s1) -- (mz3s2);
\draw[BFedge,draw=Ink] (mz3s2) -- (mz3s3);
\draw[BFedge,draw=Ink] (mz3s3) -- (z3);
\draw[BFedge,draw=Ink] (a3) -- (mu3s1);
\draw[BFedge,draw=Ink] (mu3s1) -- (mu3s2);
\draw[BFedge,draw=Ink] (mu3s2) -- (mu3s3);
\draw[BFedge,draw=Ink,opacity=.28] (mu3s3) -- (u3);
\draw[BFedge,draw=BFOutput] (v3) -- (me3s1);
\draw[BFedge,draw=BFOutput] (me3s1) -- (me3s2);
\draw[BFedge,draw=BFOutput] (me3s2) -- (me3s3);
\draw[BFedge,draw=BFOutput] (me3s3) -- (e3);
\draw[BFedge,draw=BFOutput,opacity=.28] (v3) -- (y3);
\draw[BFedge,draw=BFConnector,opacity=.28] (w1) -- (nb1);
\draw[BFedge,draw=BFConnector,opacity=.28] (nb1) -- (na1);
\draw[BFedge,draw=BFConnector,opacity=.28] (na1) -- (5.200,-2.800) -- (7.700,-2.800) -- (u2);
\draw[BFedge,draw=BFConnector] (nb1) -- (nz1s1);
\draw[BFedge,draw=BFConnector] (nz1s1) -- (nz1s2);
\draw[BFedge,draw=BFConnector] (nz1s2) -- (nz1s3);
\draw[BFedge,draw=BFConnector] (nz1s3) -- (nz1);
\draw[BFedge,draw=BFConnector,opacity=.28] (na1) -- (nu1s1);
\draw[BFedge,draw=BFConnector] (nu1s1) -- (nu1s2);
\draw[BFedge,draw=BFConnector] (nu1s2) -- (nu1s3);
\draw[BFedge,draw=BFConnector] (nu1s3) -- (nu1);
\draw[BFedge,draw=BFConnector,opacity=.28] (w2) -- (nb2);
\draw[BFedge,draw=BFConnector,opacity=.28] (nb2) -- (na2);
\draw[BFedge,draw=BFConnector,opacity=.28] (na2) -- (10.900,-2.800) -- (13.400,-2.800) -- (u3);
\draw[BFedge,draw=BFConnector,opacity=.28] (nb2) -- (nz2s1);
\draw[BFedge,draw=BFConnector] (nz2s1) -- (nz2s2);
\draw[BFedge,draw=BFConnector] (nz2s2) -- (nz2s3);
\draw[BFedge,draw=BFConnector] (nz2s3) -- (nz2);
\draw[BFedge,draw=BFConnector] (na2) -- (nu2s1);
\draw[BFedge,draw=BFConnector] (nu2s1) -- (nu2s2);
\draw[BFedge,draw=BFConnector] (nu2s2) -- (nu2s3);
\draw[BFedge,draw=BFConnector] (nu2s3) -- (nu2);
\node[BFvertex,draw=Ink,fill=Ink] at (v1) {};
\node[BFvertex,draw=Muted!50,fill=Muted!50] at (b1) {};
\node[BFvertex,draw=Ink,fill=Ink] at (a1) {};
\node[BFvertex,draw=Muted!50,fill=Muted!50] at (w1) {};
\node[BFvertex,draw=Ink,fill=Ink] at (z1) {};
\node[BFvertex,draw=Ink,fill=Ink] at (u1) {};
\node[BFsubvertex,draw=Ink,fill=Ink] at (mz1s1) {};
\node[BFsubvertex,draw=Ink,fill=Ink] at (mz1s2) {};
\node[BFsubvertex,draw=Ink,fill=Ink] at (mz1s3) {};
\node[BFsubvertex,draw=Ink,fill=Ink] at (mu1s1) {};
\node[BFsubvertex,draw=Ink,fill=Ink] at (mu1s2) {};
\node[BFsubvertex,draw=Ink,fill=Ink] at (mu1s3) {};
\node[BFvertex,draw=BFOutput,fill=BFOutput] at (e1) {};
\node[BFvertex,draw=Muted!50,fill=Muted!50] at (y1) {};
\node[BFsubvertex,draw=BFOutput,fill=BFOutput] at (me1s1) {};
\node[BFsubvertex,draw=BFOutput,fill=BFOutput] at (me1s2) {};
\node[BFsubvertex,draw=BFOutput,fill=BFOutput] at (me1s3) {};
\node[BFvertex,draw=Muted!50,fill=Muted!50] at (v2) {};
\node[BFvertex,draw=Ink,fill=Ink] at (b2) {};
\node[BFvertex,draw=Muted!50,fill=Muted!50] at (a2) {};
\node[BFvertex,draw=Ink,fill=Ink] at (w2) {};
\node[BFvertex,draw=Ink,fill=Ink] at (z2) {};
\node[BFvertex,draw=Ink,fill=Ink] at (u2) {};
\node[BFsubvertex,draw=Ink,fill=Ink] at (mz2s1) {};
\node[BFsubvertex,draw=Ink,fill=Ink] at (mz2s2) {};
\node[BFsubvertex,draw=Ink,fill=Ink] at (mz2s3) {};
\node[BFsubvertex,draw=Ink,fill=Ink] at (mu2s1) {};
\node[BFsubvertex,draw=Ink,fill=Ink] at (mu2s2) {};
\node[BFsubvertex,draw=Ink,fill=Ink] at (mu2s3) {};
\node[BFvertex,draw=BFOutput,fill=BFOutput] at (e2) {};
\node[BFvertex,draw=BFOutput,fill=BFOutput] at (y2) {};
\node[BFsubvertex,draw=BFOutput,fill=BFOutput] at (me2s1) {};
\node[BFsubvertex,draw=BFOutput,fill=BFOutput] at (me2s2) {};
\node[BFsubvertex,draw=BFOutput,fill=BFOutput] at (me2s3) {};
\node[BFvertex,draw=Ink,fill=Ink] at (v3) {};
\node[BFvertex,draw=Muted!50,fill=Muted!50] at (b3) {};
\node[BFvertex,draw=Ink,fill=Ink] at (a3) {};
\node[BFvertex,draw=Ink,fill=Ink] at (w3) {};
\node[BFvertex,draw=Ink,fill=Ink] at (z3) {};
\node[BFvertex,draw=Muted!50,fill=Muted!50] at (u3) {};
\node[BFsubvertex,draw=Ink,fill=Ink] at (mz3s1) {};
\node[BFsubvertex,draw=Ink,fill=Ink] at (mz3s2) {};
\node[BFsubvertex,draw=Ink,fill=Ink] at (mz3s3) {};
\node[BFsubvertex,draw=Ink,fill=Ink] at (mu3s1) {};
\node[BFsubvertex,draw=Ink,fill=Ink] at (mu3s2) {};
\node[BFsubvertex,draw=Ink,fill=Ink] at (mu3s3) {};
\node[BFvertex,draw=BFOutput,fill=BFOutput] at (e3) {};
\node[BFvertex,draw=Muted!50,fill=Muted!50] at (y3) {};
\node[BFsubvertex,draw=BFOutput,fill=BFOutput] at (me3s1) {};
\node[BFsubvertex,draw=BFOutput,fill=BFOutput] at (me3s2) {};
\node[BFsubvertex,draw=BFOutput,fill=BFOutput] at (me3s3) {};
\node[BFvertex,draw=BFConnector,fill=BFConnector] at (nb1) {};
\node[BFvertex,draw=Muted!50,fill=Muted!50] at (na1) {};
\node[BFvertex,draw=BFConnector,fill=BFConnector] at (nz1) {};
\node[BFvertex,draw=BFConnector,fill=BFConnector] at (nu1) {};
\node[BFsubvertex,draw=BFConnector,fill=BFConnector] at (nz1s1) {};
\node[BFsubvertex,draw=BFConnector,fill=BFConnector] at (nz1s2) {};
\node[BFsubvertex,draw=BFConnector,fill=BFConnector] at (nz1s3) {};
\node[BFsubvertex,draw=BFConnector,fill=BFConnector] at (nu1s1) {};
\node[BFsubvertex,draw=BFConnector,fill=BFConnector] at (nu1s2) {};
\node[BFsubvertex,draw=BFConnector,fill=BFConnector] at (nu1s3) {};
\node[BFvertex,draw=Muted!50,fill=Muted!50] at (nb2) {};
\node[BFvertex,draw=BFConnector,fill=BFConnector] at (na2) {};
\node[BFvertex,draw=BFConnector,fill=BFConnector] at (nz2) {};
\node[BFvertex,draw=BFConnector,fill=BFConnector] at (nu2) {};
\node[BFsubvertex,draw=BFConnector,fill=BFConnector] at (nz2s1) {};
\node[BFsubvertex,draw=BFConnector,fill=BFConnector] at (nz2s2) {};
\node[BFsubvertex,draw=BFConnector,fill=BFConnector] at (nz2s3) {};
\node[BFsubvertex,draw=BFConnector,fill=BFConnector] at (nu2s1) {};
\node[BFsubvertex,draw=BFConnector,fill=BFConnector] at (nu2s2) {};
\node[BFsubvertex,draw=BFConnector,fill=BFConnector] at (nu2s3) {};
\node[below left=3pt,text=Ink,inner sep=1pt] at (v1) {$v_1$};
\node[above=3pt,text=Muted,inner sep=1pt] at (b1) {$b_1$};
\node[above=3pt,text=Ink,inner sep=1pt] at (a1) {$a_1$};
\node[above=3pt,text=Muted,inner sep=1pt] at (w1) {$w_1$};
\node[below=3pt,text=Ink,inner sep=1pt] at (z1) {$z_1$};
\node[below=3pt,text=Ink,inner sep=1pt] at (u1) {$u_1$};
\node[above=3pt,text=BFOutput,inner sep=1pt] at (e1) {$e_1$};
\node[left=3pt,text=Muted,inner sep=1pt] at (y1) {$y_1$};
\node[below left=3pt,text=Muted,inner sep=1pt] at (v2) {$v_2$};
\node[above=3pt,text=Ink,inner sep=1pt] at (b2) {$b_2$};
\node[above=3pt,text=Muted,inner sep=1pt] at (a2) {$a_2$};
\node[above=3pt,text=Ink,inner sep=1pt] at (w2) {$w_2$};
\node[below=3pt,text=Ink,inner sep=1pt] at (z2) {$z_2$};
\node[right=5pt,text=Ink,inner sep=1pt] at (u2) {$u_2$};
\node[above=3pt,text=BFOutput,inner sep=1pt] at (e2) {$e_2$};
\node[left=10pt,text=BFOutput,inner sep=1pt] at (y2) {$y_2$};
\node[below left=3pt,text=Ink,inner sep=1pt] at (v3) {$v_3$};
\node[above=3pt,text=Muted,inner sep=1pt] at (b3) {$b_3$};
\node[above=3pt,text=Ink,inner sep=1pt] at (a3) {$a_3$};
\node[above=3pt,text=Ink,inner sep=1pt] at (w3) {$w_3$};
\node[below=3pt,text=Ink,inner sep=1pt] at (z3) {$z_3$};
\node[right=5pt,text=Muted,inner sep=1pt] at (u3) {$u_3$};
\node[above=3pt,text=BFOutput,inner sep=1pt] at (e3) {$e_3$};
\node[left=3pt,text=Muted,inner sep=1pt] at (y3) {$y_3$};
\node[above left=3pt,text=BFConnector,inner sep=1pt] at (nb1) {$b^{\prime}_1$};
\node[above right=3pt,text=Muted,inner sep=1pt] at (na1) {$a^{\prime}_1$};
\node[below=3pt,text=BFConnector,inner sep=1pt] at (nz1) {$z^{\prime}_1$};
\node[below=3pt,text=BFConnector,inner sep=1pt] at (nu1) {$u^{\prime}_1$};
\node[above left=3pt,text=Muted,inner sep=1pt] at (nb2) {$b^{\prime}_2$};
\node[above right=3pt,text=BFConnector,inner sep=1pt] at (na2) {$a^{\prime}_2$};
\node[below=3pt,text=BFConnector,inner sep=1pt] at (nz2) {$z^{\prime}_2$};
\node[below=3pt,text=BFConnector,inner sep=1pt] at (nu2) {$u^{\prime}_2$};
\node[text=Muted] at (1.600,1.35) {$M_1$};
\node[text=Muted] at (7.300,1.35) {$M_2$};
\node[text=Muted] at (13.000,1.35) {$M_3$};
\node[text=BFConnector] at (4.200,-1.53) {$N_1$};
\node[text=BFConnector] at (9.900,-1.53) {$N_2$};
\draw[Rose,line width=.85pt] ($(a2)+(-.10,-.11)$)--($(a2)+(.10,.11)$) ($(a2)+(-.10,.11)$)--($(a2)+(.10,-.11)$);
\draw[Rose,line width=.85pt] ($(b1)+(-.10,-.11)$)--($(b1)+(.10,.11)$) ($(b1)+(-.10,.11)$)--($(b1)+(.10,-.11)$);
\draw[Rose,line width=.85pt] ($(b3)+(-.10,-.11)$)--($(b3)+(.10,.11)$) ($(b3)+(-.10,.11)$)--($(b3)+(.10,-.11)$);
\draw[Rose,line width=.85pt] ($(na1)+(-.10,-.11)$)--($(na1)+(.10,.11)$) ($(na1)+(-.10,.11)$)--($(na1)+(.10,-.11)$);
\draw[Rose,line width=.85pt] ($(nb2)+(-.10,-.11)$)--($(nb2)+(.10,.11)$) ($(nb2)+(-.10,.11)$)--($(nb2)+(.10,-.11)$);
\draw[Rose,line width=.85pt] ($(u3)+(-.10,-.11)$)--($(u3)+(.10,.11)$) ($(u3)+(-.10,.11)$)--($(u3)+(.10,-.11)$);
\draw[Rose,line width=.85pt] ($(v2)+(-.10,-.11)$)--($(v2)+(.10,.11)$) ($(v2)+(-.10,.11)$)--($(v2)+(.10,-.11)$);
\draw[Rose,line width=.85pt] ($(w1)+(-.10,-.11)$)--($(w1)+(.10,.11)$) ($(w1)+(-.10,.11)$)--($(w1)+(.10,-.11)$);
\draw[Rose,line width=.85pt] ($(y1)+(-.10,-.11)$)--($(y1)+(.10,.11)$) ($(y1)+(-.10,.11)$)--($(y1)+(.10,-.11)$);
\draw[Rose,line width=.85pt] ($(y3)+(-.10,-.11)$)--($(y3)+(.10,.11)$) ($(y3)+(-.10,.11)$)--($(y3)+(.10,-.11)$);
\node[draw=BFOutput,line width=1pt,circle,minimum size=9pt,inner sep=0pt] at (y2) {};
\end{tikzpicture}
}

\newcommand{\BVCCross}[1]{%
  \node[bvc hollow] at (#1) {};
  \draw[Rose,line width=.75pt] ($(#1)+(-.13,-.13)$)--($(#1)+(.13,.13)$);
  \draw[Rose,line width=.75pt] ($(#1)+(-.13,.13)$)--($(#1)+(.13,-.13)$);
}

\newcommand{\BVCBranchFigure}[1]{%
\begin{tikzpicture}[bvc picture]
  \coordinate (bvcy1) at (2.1,4.5);
  \coordinate (bvcy2) at (4.2,4.5);
  \coordinate (bvcy3) at (6.3,4.5);
  \coordinate (bvcd1) at (10.3,4.5);
  \coordinate (bvcd2) at (12.4,4.5);
  \coordinate (bvcu1) at (2.2,1.1);
  \coordinate (bvcu2) at (3.8,1.1);
  \coordinate (bvcv1) at (7,1.1);
  \coordinate (bvcv2) at (8.6,1.1);
  \coordinate (bvcw1) at (12,1.1);
  \draw[bvc region] (.3,3.75) rectangle (14.7,6.35);
  \draw[bvc region] (.3,-.5) rectangle (14.7,2.45);
  \node[anchor=north west] at (.6,5.8) {$X$};
  \node[bvc label] at (4.2,5.95) {$Y=X\setminus D$};
  \node[bvc label] at (11.35,5.95) {$D$};
  \node[anchor=north west] at (.6,2.2) {$I$};
  \draw[bvc ghost] (bvcu1)--(bvcd1);
  \draw[bvc ghost] (bvcv2)--(bvcd2);
  \draw[bvc ghost] (bvcw1)--(bvcd1);
  \draw [bvc edge] (bvcy1)--(bvcy2);
  \draw [bvc edge] (bvcy2)--(bvcy3);
  \draw [bvc ghost] (bvcd1)--(bvcd2);
  \foreach \u in {bvcu1,bvcu2}{
    \draw[bvc edge] (\u)--(bvcy1);
    \draw[bvc edge] (\u)--(bvcy2);
  }
  \foreach \v in {bvcv1,bvcv2}\draw[bvc edge] (\v)--(bvcy3);
  \ifnum#1=1\relax
    \draw[bvc bucket] (1.55,.53) rectangle (4.45,1.62);
    \draw[bvc bucket] (6.35,.53) rectangle (9.25,1.62);
    \draw[bvc bucket] (11.25,.53) rectangle (12.75,1.62);
    \node[bvc label] at (3,-.05) {$B_{\{y_1,y_2\}}$};
    \node[bvc label] at (7.8,-.05) {$B_{\{y_3\}}$};
    \node[bvc label] at (12,-.05) {$B_{\emptyset}$};
  \fi
  \foreach \i in {1,2,3}{
    \node[bvc survivor] at (bvcy\i) {};
    \node[above=4pt] at (bvcy\i) {$y_\i$};
  }
  \foreach \i in {1,2}{
    \BVCCross{bvcd\i}
    \node[above=4pt] at (bvcd\i) {$d_\i$};
  }
  \foreach \p/\lab in {bvcu1/u_1,bvcu2/u_2,bvcv1/v_1,bvcv2/v_2,bvcw1/w}{
    \node[bvc vertex] at (\p) {};

  }
\end{tikzpicture}%
}

\newcommand{\GlobalGraphPaper}[1][0]{%
\begin{tikzpicture}[x=.82cm,y=.65cm,font=\small,
  ggv label/.style={font=\small,inner sep=1pt},
  ggv attach/.style={draw=Muted!62,line width=.7pt,line cap=round}]
\path[use as bounding box] (0,-.53) rectangle (16.4,6.27);
\node[font=\small] at (8.2,6.0)
  {$k=3,\quad E(\Gamma)=\{e\},\quad e=\{(1,2),(3,1)\}$};
\foreach \i/\yy in {1/5.12,2/3.69,3/2.26}{
  \coordinate (ggvr\i) at (2,\yy);
  \coordinate (ggvc\i) at (14.4,\yy);
  \ifnum#1=0
    \draw[pathseg,Muted!80] (.35,\yy)--(ggvr\i);
    \draw[pathseg,Muted!80] (ggvc\i)--(16.05,\yy);
    \node[hollow] at (.35,\yy) {};
    \node[hollow] at (16.05,\yy) {};
    \node[ggv label] at (1.12,\yy+.24) {$L$};
    \node[ggv label] at (15.23,\yy+.24) {$L$};
  \else
    \node at (.9,\yy) {$P_L$};\node at (15.5,\yy) {$P_L$};
  \fi
}

\foreach \i/\yy in {1/5.12,2/3.69,3/2.26}{
  \filldraw[fill=Soft,draw=Muted!60,line width=.7pt]
    (3.95,\yy-.68) rectangle (11.7,\yy+.43);
  \node[ggv label] at (4.55,\yy-.30) {$T_{\i}^{\rm row}$};
  \foreach \j/\xx in {1/5.75,2/8.2,3/10.65}{
    \coordinate (ggvy\i\j) at (\xx,\yy);
  }
}
\filldraw[fill=Soft,draw=Muted!60,line width=.7pt]
  (3.95,.15) rectangle (11.7,1.26);
\node[ggv label] at (4.3,.53) {$T^e$};
\foreach \n/\xx in {1/5.3,2/7.2,3/9.1,4/11.0}{
  \coordinate (ggvx\n) at (\xx,.83);
}

\ifnum#1=0
  \foreach \i/\yy in {1/5.12,2/3.69,3/2.26}{
    \foreach \j/\xx/\cy in {1/5.75/5.12,2/8.2/3.69,3/10.65/2.26}{
      \draw[ggv attach] (ggvr\i)
        .. controls (4.0,\yy+.35) and (\xx-.45,\yy+.35) .. (ggvy\i\j);
      \draw[ggv attach] (ggvy\i\j)
        .. controls (\xx+.45,\yy+.35) and (11.9,\yy+.35) .. (12+.2*\j,\yy+.20+.05*\j)
        .. controls (13.2,\yy+.35) and (13.65,\cy) .. (ggvc\j);
    }
  }
  \foreach \n/\xx/\a/\b/\ry/\cy in {1/5.3/1/1/5.12/5.12,2/7.2/1/3/5.12/2.26,3/9.1/3/2/2.26/3.69,4/11.0/3/3/2.26/2.26}{
    \draw[ggv attach] (ggvr\a)
      .. controls (3.0,\ry) and (3.2,1.18) .. (3.6+.15*\n,1.18)
      .. controls (4.5,1.18) and (\xx-.4,1.18) .. (ggvx\n);
    \draw[ggv attach] (ggvx\n)
      .. controls (\xx+.4,1.18) and (11.9,1.18) .. (12+.15*\n,1.18)
      .. controls (13.2,1.18) and (13.6,\cy) .. (ggvc\b);
  }

  \draw[ed,Teal,line width=1pt] (ggvr2)
    .. controls (4,4.04) and (7.75,4.04) .. (ggvy22);
  \draw[ed,Teal,line width=1pt] (ggvy22)
    .. controls (8.65,4.04) and (11.9,4.04) .. (12.4,3.99)
    .. controls (13.2,4.04) and (13.65,3.69) .. (ggvc2);
\fi

\foreach \i/\yy in {1/5.12,2/3.69,3/2.26}{
  \ifnum#1=0
    \node[vertex] at (ggvr\i) {};
    \node[vertex] at (ggvc\i) {};
  \else
    \foreach \p in {ggvr\i,ggvc\i}{
      \node[vertex,fill=Muted!25,draw=Muted!25] at (\p) {};
      \draw[Rose,line width=1.25pt] ($(\p)+(-.11,-.11)$)--($(\p)+(.11,.11)$)
        ($(\p)+(-.11,.11)$)--($(\p)+(.11,-.11)$);
    }
  \fi
  \node[ggv label,anchor=north east] at (1.94,\yy-.12) {$r_\i$};
  \node[ggv label,anchor=north west] at (14.46,\yy-.12) {$c_\i$};
  \foreach \j/\xx in {1/5.75,2/8.2,3/10.65}{
    \node[output] at (ggvy\i\j) {};
    \node[ggv label,text=Teal,anchor=north] at (\xx,\yy-.12) {$(\i,\j)$};
  }
}
\foreach \n/\xx/\a/\b in {1/5.3/1/1,2/7.2/1/3,3/9.1/3/2,4/11.0/3/3}{
  \node[output] at (ggvx\n) {};
  \node[ggv label,text=Teal,anchor=north] at (\xx,.71) {$(\a,\b)$};
}
\ifnum#1=0
  \node[text=Muted,font=\small] at (8.2,-.25)
    {};
\else
  \node[text=Teal,font=\small] at (8.2,-.25)
    {Deleting $R\cup C$ leaves four selector trees and six copies of $P_L$.};
\fi
\end{tikzpicture}%
}

\newcommand{\CorrectnessPictures}{%
\begin{tikzpicture}[
  y=0.75cm,
  vertex/.style={
    circle,
    fill=black,
    inner sep=0pt,
    minimum size=3.5pt
  },
  every label/.style={font=\footnotesize},
  every node/.style={font=\footnotesize},
  line width=0.55pt
]

    \begin{scope}
      \node[vertex] (a-left) at (-0.9,0) {};
      \node[vertex,label=above:{$c_j$}]
        (a-cj) at (0,0) {};
      \node[vertex,
        label=above:{$x$},
        label=below:{$(i,j)$}]
        (a-x) at (0.95,0) {};
      \node[vertex,label=above:{$r_i$}]
        (a-ri) at (1.9,0) {};
      \node[vertex,
        label=above:{$y$},
        label=below:{$(i,j')$}]
        (a-y) at (2.85,0) {};
      \node[vertex,label=above:{$c_{j'}$}]
        (a-cjp) at (3.8,0) {};
      \node[vertex] (a-right) at (4.7,0) {};

      \draw (a-cj) -- (a-x) -- (a-ri) -- (a-y) -- (a-cjp);
      \draw[dashed] (a-left) -- node[above=3pt] {$L$} (a-cj);
      \draw[dashed] (a-cjp) -- node[above=3pt] {$L$} (a-right);

      \node at (1.9,-0.85) {$j\ne j'$};
      \node at (1.9,-1.65) {(a)};
    \end{scope}

    \begin{scope}[xshift=6.7cm]
      \node[vertex] (b-left) at (-1.5,0) {};
      \node[vertex,label=above left:{$r_i$}]
        (b-r) at (0,0) {};
      \node[vertex,label=above right:{$c_j$}]
        (b-c) at (3.2,0) {};
      \node[vertex] (b-right) at (4.7,0) {};

      \node[vertex,label=above:{$(i,j)$}]
        (b-top) at (1.6,1) {};
      \node[vertex,label=above:{$(i,j)$}]
        (b-middle) at (1.6,0) {};
      \node[vertex,label=above:{$(i,j)$}]
        (b-bottom) at (1.6,-1) {};

      \draw (b-r) -- (b-top) -- (b-c);
      \draw (b-r) -- (b-middle) -- (b-c);
      \draw (b-r) -- (b-bottom) -- (b-c);

      \draw[dashed] (b-left) -- node[above=3pt] {$L$} (b-r);
      \draw[dashed] (b-c) -- node[above=3pt] {$L$} (b-right);

      \node at (1.6,-1.65) {(b)};
    \end{scope}

  \end{tikzpicture}%
}

\newcommand{\TreePathPackingPicture}[1]{%
  \begin{tikzpicture}[
    x=.80cm,
    y=.60cm,
    font=\footnotesize,
    background edge/.style={
      draw=black!22,
      line width=.55pt
    },
    selected path/.style={
      line width=1.8pt,
      line cap=round,
      line join=round
    },
    path A/.style={selected path,draw=blue!70!black},
    path B/.style={selected path,draw=orange!85!black},
    path C/.style={selected path,draw=teal!80!black},
    path D/.style={selected path,draw=violet!80!black},
    every node/.style={inner sep=1pt}
  ]
    \path[use as bounding box]
      (-1.1,-4.25) rectangle (15.1,2.5);

    \foreach \i in {1,2,3}{
      \pgfmathsetmacro{\s}{5.7*(\i-1)}

      \coordinate (v\i) at (\s,0);
      \coordinate (b\i) at ({\s+1},0);
      \coordinate (a\i) at ({\s+2},0);
      \coordinate (w\i) at ({\s+3},0);
      \coordinate (z\i) at ({\s+1},-1.55);
      \coordinate (u\i) at ({\s+2},-1.55);
      \coordinate (e\i) at (\s,1.9);
      \coordinate (y\i) at ({\s-.55},.7);

      \draw[background edge]
        (v\i) -- (b\i) -- (a\i) -- (w\i);
      \draw[background edge] (b\i) -- (z\i);
      \draw[background edge] (a\i) -- (u\i);
      \draw[background edge] (v\i) -- (e\i);

      \ifnum#1=1\relax
        \draw[background edge] (v\i) -- (y\i);
      \fi
    }

    \foreach \i in {1,2}{
      \pgfmathsetmacro{\s}{5.7*(\i-1)+3.7}
      \pgfmathtruncatemacro{\next}{\i+1}

      \coordinate (nb\i) at (\s,-2.2);
      \coordinate (na\i) at ({\s+1},-2.2);
      \coordinate (nz\i) at (\s,-3.7);
      \coordinate (nu\i) at ({\s+1},-3.7);

      \draw[background edge]
        (w\i) -- (nb\i) -- (na\i);
      \draw[background edge]
        (na\i)
        -- ({\s+1.5},-2.8)
        -- ({\s+4},-2.8)
        -- (u\next);
      \draw[background edge] (nb\i) -- (nz\i);
      \draw[background edge] (na\i) -- (nu\i);
    }

    \ifnum#1=1\relax

      \foreach \i in {1,2,3}{

        \draw[path A] (a\i) -- (b\i) -- (z\i);

        \draw[path B] (e\i) -- (v\i) -- (y\i);
      }

      \foreach \i in {1,2}{
        \pgfmathsetmacro{\s}{5.7*(\i-1)+3.7}
        \pgfmathtruncatemacro{\next}{\i+1}

        \draw[path C] (w\i) -- (nb\i) -- (nz\i);

        \draw[path D]
          (nu\i) -- (na\i)
          -- ({\s+1.5},-2.8)
          -- ({\s+4},-2.8)
          -- (u\next);
      }

    \else

      \foreach \i in {1,2,3}{

        \draw[path A] (u\i) -- (a\i) -- (w\i);

        \draw[path C] (z\i) -- (b\i) -- (v\i);
      }

      \foreach \i in {1,2}{

        \draw[path B] (nz\i) -- (nb\i) -- (na\i);
      }

    \fi

    \foreach \i in {1,2,3}{
      \pgfmathsetmacro{\s}{5.7*(\i-1)}

      \foreach \p in {v,b,a,w,z,u,e}{
        \fill (\p\i) circle[radius=1.5pt];
      }

      \ifnum#1=1\relax
        \fill (y\i) circle[radius=1.5pt];
      \fi

      \foreach \j in {1,2,3}{
        \fill ({\s+1},{-1.55*\j/4})
          circle[radius=1.5pt];
        \fill ({\s+2},{-1.55*\j/4})
          circle[radius=1.5pt];
        \fill (\s,{1.9*\j/4})
          circle[radius=1.5pt];
      }

      \node[below left=3pt] at (v\i) {$v_{\i}$};
      \node[above=3pt] at (b\i) {$b_{\i}$};
      \node[above=3pt] at (a\i) {$a_{\i}$};
      \node[above=3pt] at (w\i) {$w_{\i}$};
      \node[below=3pt] at (z\i) {$z_{\i}$};
      \node[above=3pt] at (e\i) {$e_{\i}$};

      \ifnum\i=1\relax
        \node[below=3pt] at (u\i) {$u_{\i}$};
      \else
        \node[right=4pt] at (u\i) {$u_{\i}$};
      \fi

      \ifnum#1=1\relax
        \node[left=3pt] at (y\i) {$y_{\i}$};
      \fi

      \node[text=black!55] at ({\s+1.6},1.35)
        {$M_{\i}$};
    }

    \foreach \i in {1,2}{
      \pgfmathsetmacro{\s}{5.7*(\i-1)+3.7}

      \foreach \p in {nb,na,nz,nu}{
        \fill (\p\i) circle[radius=1.5pt];
      }

      \foreach \j in {1,2,3}{
        \fill (\s,{-2.2-1.5*\j/4})
          circle[radius=1.5pt];
        \fill ({\s+1},{-2.2-1.5*\j/4})
          circle[radius=1.5pt];
      }

      \node[above left=3pt] at (nb\i) {$b_{\i}'$};
      \node[above right=3pt] at (na\i) {$a_{\i}'$};
      \node[below=3pt] at (nz\i) {$z_{\i}'$};
      \node[below=3pt] at (nu\i) {$u_{\i}'$};

      \node[text=black!55] at ({\s+.5},-1.5)
        {$N_{\i}$};
    }
  \end{tikzpicture}%
}

\newcommand{\PackingWithOutputsPicture}{%
  \TreePathPackingPicture{1}%
}

\newcommand{\PackingWithoutOutputsPicture}{%
  \TreePathPackingPicture{2}%
}

\newcommand{\RepeatedOutputsPfivePicture}{%
  \begin{tikzpicture}[
    vertex/.style={
      circle,
      fill=black,
      inner sep=0pt,
      minimum size=4pt
    },
    every node/.style={font=\small},
    every label/.style={font=\small},
    line width=.6pt
  ]
    \node[vertex,label=left:{$r_i$}]
      (pf-r) at (0,0) {};
    \node[vertex,label=right:{$c_j$}]
      (pf-c) at (3.6,0) {};

    \node[vertex,label=above:{$x_1$}]
      (pf-x1) at (1.8,1.2) {};
    \node[vertex,label=above:{$x_2$}]
      (pf-x2) at (1.8,0) {};
    \node[vertex,label=below:{$x_3$}]
      (pf-x3) at (1.8,-1.2) {};

    \draw[draw=black!35] (pf-x1) -- (pf-c);
    \draw[draw=black!35] (pf-r) -- (pf-x3);

    \draw[draw=blue!70!black,line width=1.5pt]
      (pf-x1) -- (pf-r) -- (pf-x2) -- (pf-c) -- (pf-x3);
  \end{tikzpicture}%
}
\crefname{observation}{Observation}{Observations}
\Crefname{observation}{Observation}{Observations}
\tikzset{
  proof box/.style={draw=black!45,fill=white,line width=.5pt,
    align=center,inner xsep=4pt,inner ysep=4pt,font=\small},
  proof arrow/.style={-{Latex[length=1.8mm]},draw=black!65,line width=.6pt}
}

\newcommand{\VCProofMap}{%
\begin{tikzpicture}[x=1cm,y=1cm]
  \node[proof box,text width=3.6cm] (bounds) at (0,0)
    {\Cref{lem:vc-high-deg-bound,lem:vc-type-bound}};
  \node[proof box,text width=3.6cm] (exchange) at (4.7,0)
    {\Cref{lem:vc-representatives}};
  \node[proof box,text width=3.6cm] (twins) at (9.4,0)
    {\Cref{lem:vc-false-twin-reduce}};
  \node[proof box,text width=3.6cm] (size) at (0,-1.15)
    {$|V(G')| = O_{\F}(t)$};
  \node[proof box,text width=3.6cm] (complete) at (4.7,-1.15)
    {Completeness\\\Cref{lem:vc-solution-represented}};
  \node[proof box,text width=3.6cm] (sound) at (9.4,-1.15)
    {Soundness\\\Cref{lem:vc-compression}};
  \node[proof box,text width=4.2cm,fill=black!3] (result) at (4.7,-2.45)
    {\Cref{thm:main-vc}\\$2^{O_{\F}(t)}n^{O(1)}$};
  \draw[proof arrow] (bounds)--(size);
  \draw[proof arrow] (exchange)--(complete);
  \draw[proof arrow] (twins)--(sound);
  \draw[proof arrow] (size.south) |- (result.west);
  \draw[proof arrow] (complete)--(result);
  \draw[proof arrow] (sound.south) |- (result.east);
\end{tikzpicture}%
}
\newcommand{\FVSProofMap}{%
  \begin{tikzpicture}[
    box/.style={
      draw,
      rounded corners=1pt,
      align=center,
      font=\footnotesize,
      inner sep=4pt,
      text width=4.1cm,
      minimum height=0.75cm
    },
    arr/.style={-{Latex[length=1.6mm]}, semithick}
  ]
    \node[box,text width=5.4cm] (tree) at (0,0)
      {Properties of $T_{h,q}$\\
       \Cref{lem:tree-properties}};

    \node[box] (counts) at (-2.7,-1.65)
      {Exactly $4q-2$ deletions per copy;\\
       at least one output remains};

    \node[box] (choice) at (2.7,-1.65)
      {Choose one output per copy\\
       and isolate it inside the tree};

    \node[box] (independent) at (-2.7,-3.3)
      {Remaining outputs determine\\
       a permutation independent set};

    \node[box] (free) at (2.7,-3.3)
      {The remaining components\\
       contain no $P_h$ minor};

    \node[box,text width=8.5cm] (equivalence) at (0,-5)
      {$\Gamma$ has a permutation independent set
       if and only if
       $G$ has a $P_h$-deletion set of size at most $B$};

\draw[arr] (tree.south) -- (counts.north)
  node[midway, left=16pt, font=\footnotesize] {Soundness};
\draw[arr] (tree.south) -- (choice.north)
  node[midway, right=9pt, font=\footnotesize] {Completeness};
    \draw[arr] (counts.south) -- (independent.north);
    \draw[arr] (choice.south) -- (free.north);
\draw[arr] (independent.south)
  -- (independent.south |- equivalence.north);
\draw[arr] (free.south)
  -- (free.south |- equivalence.north);
  \end{tikzpicture}%
}

\makeatletter
\newcommand{\primaryfootnote}[1]{%
  \ifthmt@thisistheone
    \footnote{#1}%
  \fi
}
\makeatother

\newcommand{\F}{\mathcal{F}}
\newcommand{\tw}{\operatorname{tw}}
\newcommand{\td}{\operatorname{td}}

\newcommand{\vc}{\operatorname{vc}}
\newcommand{\fvs}{\operatorname{fvs}}
\newcommand{\problemname}{$\F$\textsc{-Minor-Free Deletion}}
\newcommand{\permset}{$k\times k$\textsc{-Permutation Independent Set}}

\newcommand{\Yes}{\textsc{Yes}}
\newcommand{\No}{\textsc{No}}
\newcommand{\Types}{\mathcal{N}_Y}
\newcommand{\Type}{N_Y}

\begin{document}

\maketitle

\begin{abstract}
For a fixed finite collection of graphs $\mathcal{F}$, the
\textsc{$\mathcal{F}$-Minor-Free Deletion} problem asks, given a graph $G$
and an integer $k$, whether there is a set $S \subseteq V(G)$ with
$|S| \leq k$ such that $G - S$ contains no graph of $\mathcal{F}$ as a minor.
This generalizes \textsc{Vertex Cover}, \textsc{Feedback Vertex Set},
and \textsc{Vertex Planarization}, among many others.
Baste, Sau, and
Thilikos [JCSS 2020, SICOMP 2023] showed that, assuming the Exponential Time
Hypothesis (ETH), there are collections $\mathcal{F}$ for which the problem
cannot be solved in time $2^{o(\mathrm{tw}\log\mathrm{tw})}\cdot n^{O(1)}$
on $n$-vertex graphs of treewidth $\mathrm{tw}$, and they gave an algorithm
that matches this bound.

We ask whether this super-exponential dependence persists under
more restrictive structural parameters. First, we strengthen the lower bound
of Baste, Sau, and Thilikos. Assuming the ETH, for every fixed $h \geq 6$,
\textsc{$\{P_h\}$-Minor-Free Deletion}, where $P_h$ is the path on $h$ vertices,
cannot be solved in time
$2^{o(t\log t)}\cdot n^{O(1)}$, where
$t=\mathrm{fvs}+\mathrm{td}$ is the sum of the feedback vertex set number
and treedepth of the input graph.
These two parameters $\mathrm{fvs}$ and $\mathrm{td}$ are incomparable with each other,
but both individually upper bound the treewidth of the graph.
Together with the $2^{O(\mathrm{tw}\log\mathrm{tw})}\cdot n^{O(1)}$ time algorithm by Baste, Sau, and Thilikos [SICOMP 2023], our result shows that our lower bound is tight.

In contrast to the lower bound for the parameter $\mathrm{fvs} + \mathrm{td}$, we show that asymptotically faster algorithms are possible when the parameter is the vertex cover number $\mathrm{vc}$ of the input graph.
In particular we show that, for any finite family $\mathcal{F}$, the
\textsc{$\mathcal{F}$-Minor-Free Deletion} problem can be solved in time
$2^{O(\mathrm{vc})}\cdot n^{O(1)}$.
Taken together, our results show that the
boundary between single-exponential and super-exponential complexity
for this problem lies between the vertex cover number and the
feedback vertex set number (plus treedepth).
\end{abstract}
\newpage
\section{Introduction}\label{sec:intro}
Many classical graph problems can be phrased as \emph{vertex deletion
problems}: given a graph~$G$ and an integer~$k$, delete at most $k$ vertices
so that the resulting graph belongs to a prescribed graph class~$\mathcal{C}$.
When $\mathcal{C}$ is closed under taking minors, the Graph Minor Theorem of
Robertson and Seymour~\cite{RobertsonS2004Graph} guarantees that $\mathcal{C}$ is
characterized by a finite collection~$\mathcal{F}$ of forbidden minors. This
leads to the following (meta) problem.\footnote{We use the decision version throughout the paper. Binary search on $k$ gives the optimization version with only a logarithmic overhead.}

\begin{center}
\fbox{\parbox{0.92\linewidth}{
\textsc{$\mathcal{F}$-Minor-Free Deletion}\\
\emph{Input:} A graph $G$ and a nonnegative integer $k$.\\
\emph{Question:} Is there a set $S \subseteq V(G)$ with $|S| \le k$ such that
$G - S$ contains no graph of $\mathcal{F}$ as a minor?
}}
\end{center}

This framework captures a broad range of well-studied NP-hard vertex-deletion graphs problems. For instance,
$\mathcal{F}=\{K_2\}$ yields \textsc{Vertex Cover}, $\mathcal{F}=\{K_3\}$
yields \textsc{Feedback Vertex Set}, and $\mathcal{F}=\{K_5,K_{3,3}\}$ yields
\textsc{Vertex Planarization}.
Other choices of $\mathcal{F}$ yield vertex-deletion to bounded genus graphs, to bounded treewidth/pathwidth/treedepth graphs, and many more.
Consequently, \textsc{$\mathcal{F}$-Minor-Free
Deletion} has been a central object of study in parameterized complexity, with
a rich body of work on algorithms, kernels, and approximation for various
families~$\mathcal{F}$. See for
instance~\cite{DBLP:conf/focs/FominLMS12,DBLP:journals/talg/0002LPRRSS16,DBLP:conf/soda/JansenLS14,DBLP:journals/algorithmica/KociumakaP19}.

\subparagraph*{Parameterization by treewidth.}
A particularly fruitful line of research considers structural parameters of
the input graph. Since \textsc{$\mathcal{F}$-Minor-Free Deletion} is
expressible in monadic second-order logic, Courcelle's
Theorem~\cite{Courcelle90} implies that it can be solved in time
$f_{\mathcal{F}}(\tw)\cdot n^{O(1)}$ on $n$-vertex graphs of
treewidth~$\tw$, for some computable function~$f_{\mathcal{F}}$. However, the
function obtained in this way is astronomically large, which raises the
fine-grained question of determining the \emph{smallest} possible
function~$f_{\mathcal{F}}$ for each collection~$\mathcal{F}$.

The question, we address in this work, is closely tied to a broader theme in parameterized complexity:
distinguishing problems solvable in \emph{single-exponential} time
$2^{O(\tw)}\cdot n^{O(1)}$ from those requiring \emph{slightly
superexponential} time $2^{\Theta(\tw\log\tw)}\cdot n^{O(1)}$. On the
algorithmic side, the Cut\&Count technique~\cite{DBLP:journals/talg/CyganNPPRW22} and the
rank-based approach~\cite{DBLP:journals/iandc/BodlaenderCKN15} yield single-exponential
algorithms, parameterized by treewidth, for many connectivity problems, including \textsc{Feedback Vertex
Set}.
On the lower-bound side, Lokshtanov, Marx, and
Saurabh~\cite{LokshtanovMS2018Slightly} developed a framework for ruling out
$2^{o(k\log k)}\cdot n^{O(1)}$-time algorithms under the Exponential Time
Hypothesis (ETH)~\cite{DBLP:journals/jcss/ImpagliazzoP01,DBLP:journals/jcss/ImpagliazzoPZ01}, which was
subsequently applied to deletion problems parameterized by treewidth by
Pilipczuk~\cite{DBLP:journals/dam/Pilipczuk17} for \textsc{Vertex Planarization} and by Bonnet
et al.~\cite{DBLP:journals/algorithmica/BonnetBKM19} for generalized feedback vertex set problems.

In a series of articles, Baste, Sau, and
Thilikos~\cite{DBLP:journals/siamdm/BasteST20,BasteST2020Single,BasteST2020Hitting,BasteST2023Hitting}
carried out this programme for \textsc{$\mathcal{F}$-Minor-Free Deletion}.
They gave an algorithm running in time
$2^{O(\tw\log\tw)}\cdot n^{O(1)}$ for every fixed finite
collection~$\mathcal{F}$~\cite{BasteST2023Hitting}, and they proved that, under
the ETH, this is optimal for many collections: in particular, if
$\mathcal{F}=\{H\}$ where $H$ is a connected graph that is either $P_5$ or not
a minor of the banner\footnote{The \emph{banner} is the graph obtained from
the cycle $C_4$ by adding a pendant edge, that is, a new vertex adjacent to
exactly one vertex of the cycle.}, then the problem cannot be solved in time
$2^{o(\tw\log\tw)}\cdot n^{O(1)}$~\cite{BasteST2020Hitting}. Together with
single-exponential algorithms for the remaining
cases~\cite{BasteST2020Single}, this yields a complete dichotomy for a single
connected forbidden minor~$H$: the problem is solvable in time
$2^{\Theta(\tw)}\cdot n^{O(1)}$ if $H$ is a contraction of the chair\footnote{The \emph{chair} is the graph
obtained from the claw $K_{1,3}$ by subdividing one of its edges once. It is
also known as the \emph{fork}.} or the
banner, and requires time $2^{\Theta(\tw\log\tw)}\cdot n^{O(1)}$ otherwise,
assuming the ETH.

The same dichotomy also holds for treedepth: the upper bounds follow
from the fact that the treewidth of any graph is upper bounded by its treedepth minus 1 $\tw(G)\leq \td(G) -1$, and the lower bounds of
\cite{BasteST2020Hitting} are explicitly observed there to hold for treedepth as well.
This follows an earlier treedepth lower bound of Pilipczuk for \textsc{Vertex Planarization}~\cite{DBLP:journals/dam/Pilipczuk17}.

\subparagraph*{Parameterizing by feedback vertex set and vertex cover number.}
Treewidth is, however, only one point in a hierarchy of structural
parameters. Two prominent parameters that upper bound treewidth are the
\emph{feedback vertex set number}~$\fvs$ and the \emph{vertex cover
number}~$\vc$ of the input graph, which satisfy
$  \tw \le \fvs + 1 \le \vc + 1$.

Since these parameters are larger, a
lower bound in terms of them is \emph{stronger} than the corresponding bound
in terms of treewidth, while an algorithm parameterized by them may be asymptotically \emph{better} than what is possible for treewidth. This motivates the
following question:

\begin{quote}
\emph{Does the $2^{\Omega(\tw\log\tw)}$ barrier in the running time for
\textsc{$\mathcal{F}$-Minor-Free Deletion} persist under the feedback vertex
set number and/or the vertex cover number parameterization?}
\end{quote}

\noindent
We answer this question for both parameters by showing that the super-exponential barrier persists for $\fvs$,
but disappears for $\vc$.

Parameterizations by $\fvs$ and $\vc$, and more generally by the size of a
modulator to a simple graph class, have been studied extensively for
\textsc{$\mathcal{F}$-Minor-Free Deletion} itself, mostly from the
kernelization point of view.
Fomin, Jansen, and Pilipczuk~\cite{FominJP2014Preprocessing}
studied when a small vertex cover helps for minor and subgraph problems.
For families $\mathcal{F}$ consisting of connected graphs,
Jansen and Pieterse~\cite{DBLP:journals/tcs/JansenP20}
gave polynomial kernels parameterized by the size of a
treedepth-$\eta$ modulator for every fixed $\eta$,
which in particular covers the vertex cover number.
Donkers and Jansen~\cite{DonkersJ2021Turing} obtained a Turing
kernelization dichotomy for structural parameterizations, and Bougeret,
Brandwein, and Sau~\cite{DBLP:conf/stacs/BougeretBS26} recently obtained kernelization
dichotomies with respect to the vertex-deletion distance to minor-closed
classes. On the algorithmic side, hybrid parameters combining a modulator
with treewidth or elimination distance were studied by Eiben, Ganian, Hamm,
and Kwon~\cite{DBLP:journals/jcss/EibenGHK21} and by Jansen, de Kroon, and
W{\l}odarczyk~\cite{DBLP:conf/stoc/JansenK021}. In contrast, the fine-grained dependence of
the running time on such parameters has received little attention in the past and in this work we make a step in that direction.

\subsection{Our contributions}
\label{sec:contributions}

A graph contains $P_h$, the path
on $h$ vertices, as a minor if and only if it contains $P_h$ as a subgraph.
Hence \textsc{$\{P_h\}$-Minor-Free Deletion} asks to delete at most $k$
vertices so that the remaining graph has no path on $h$ vertices; this
problem has also been studied under the name \textsc{$h$-Path Vertex
Cover}~\cite{DBLP:journals/dam/BresarKKS11}.
Our first result shows that already
\textsc{$\{P_h\}$-Minor-Free Deletion} does not admit
a single-exponential algorithm parameterized by the sum
of the feedback vertex set number and treedepth, unless ETH fails.

\begin{restatable}{theorem}{mainlowerbound}\label{thm:lower}
For every fixed integer $h\ge 6$, unless ETH fails,
\textsc{$\{P_h\}$-Minor-Free Deletion} cannot be solved in time
$2^{o(t\log t)}\cdot n^{O(1)}$,
where $n=|V(G)|$ and $t=\fvs(G)+\td(G)$,
with $\fvs(G)$ denoting the feedback vertex set number of $G$
and $\td(G)$ its treedepth.
This lower bound holds even when a minimum feedback vertex set
of $G$ is given as part of the input.
\end{restatable}

Theorem~\ref{thm:lower} is proved in Section~\ref{sec:fvs-lower-bound}.
Feedback vertex set number and treedepth are incomparable parameters: paths have feedback vertex set number zero and unbounded treedepth, whereas disjoint unions of triangles have treedepth three and unbounded feedback vertex set number.
\Cref{thm:lower} gives a lower bound for their sum, which is stronger than a lower bound for either parameter alone.
In particular, our result shows that even when both parameters are simultaneously bounded, this is not enough to obtain a single-exponential algorithm under ETH.

\begin{corollary}\label{cor:lower-fvs}
For every fixed integer $h\ge 6$, unless ETH fails,
\textsc{$\{P_h\}$-Minor-Free Deletion} cannot be solved in time
$2^{o(\fvs(G)\log\fvs(G))}\cdot n^{O(1)}$.
This lower bound holds even when a minimum feedback vertex set
of $G$ is given as part of the input.
\end{corollary}

\begin{corollary}\label{cor:lower-td}
For every fixed integer $h\ge 6$, unless ETH fails,
\textsc{$\{P_h\}$-Minor-Free Deletion} cannot be solved in time
$2^{o(\td(G)\log\td(G))}\cdot n^{O(1)}$.
This lower bound holds even when a minimum feedback vertex set
of $G$ is given as part of the input.
\end{corollary}

\Cref{cor:lower-td} recovers the treedepth lower bound
of Baste, Sau, and Thilikos~\cite{BasteST2020Hitting}
for $\mathcal{F}=\{P_h\}$ with $h\ge 6$.
Since $\tw(G)\le\fvs(G)+1$ and $\tw(G) \leq \td(G) -1$,
their $2^{O(\tw\log\tw)}\cdot n^{O(1)}$-time
algorithm~\cite{BasteST2023Hitting} gives matching upper bounds
for both parameters.
Thus, \Cref{cor:lower-fvs,cor:lower-td} are tight under ETH.
For $1\leq h\leq 4$,
\textsc{$\{P_h\}$-Minor-Free Deletion} is solvable in
single-exponential time with respect to
treewidth~\cite{BasteST2020Single,BasteST2023Hitting},
and hence with respect to $\fvs$.
Our result therefore leaves \textsc{$\{P_5\}$-Minor-Free Deletion} as a {$\mathcal{F}$-Minor-Free Deletion} problem whose
tight parameterized complexity with respect to the parameter $\fvs$ remains open.

Our second result shows that the super-exponential dependence is no longer needed, for any family $\mathcal{F}$,
when we
parameterize the problem by the vertex cover number.

\begin{restatable}{theorem}{mainupperbound}\label{thm:upper}
    For every fixed finite collection $\mathcal{F}$ of graphs,
\textsc{$\mathcal{F}$-Minor-Free Deletion} can be solved in time
$2^{O_{\mathcal{F}}(\vc)}\cdot n^{O(1)}$ on $n$-vertex graphs, where $\vc$ is the vertex
cover number of the input graph.\primaryfootnote{The constant hidden in $O_{\mathcal F}(\vc)$ may depend on $\mathcal F$, whereas the exponent hidden in $n^{O(1)}$ is independent of $\mathcal F$.}
\end{restatable}

 Theorem~\ref{thm:upper} is proved in Section~\ref{sec:vc-algo}.
Note that \cref{thm:upper} does not require a vertex cover to be given
with the input, since a minimum vertex cover can be computed in time
$1.25284^{\vc}\cdot n^{O(1)}$ by the work of Harris and Narayanaswamy~\cite{DBLP:conf/stacs/0001N24}.

\subsection{Overview of Our Techniques}
\label{sec:techniques}

We use the following problem as the source of our reduction.

\begin{center}
\fbox{\parbox{0.92\linewidth}{
\permset\\
\emph{Input:} A graph $\Gamma$ with vertex set $[k]\times[k]$.\\
\emph{Question:} Is there a permutation $\pi:[k]\to[k]$ such that
$\{(i,\pi(i)) : i\in[k]\}$ is an independent set of $\Gamma$?
}}
\end{center}

\begin{lemma}[\cite{LokshtanovMS2018Slightly}]\label{lem:permutation-eth}
  {\permset} has no
  $2^{o(k\log k)}k^{O(1)}$ algorithm unless the ETH fails.
\end{lemma}

\subparagraph*{Lower bound.}
Our reduction uses two types of tree-shaped gadgets: a row gadget chooses a pair $(i,j)$ representing $\pi(i)=j$, while an edge gadget ensures that the two endpoints of an edge of $\Gamma$ are not both chosen.
The gadget is obtained by modifying the clause gadget of Donkers and Jansen~\cite{DonkersJ2021Turing} for $P_h$.
They used the gadget to show that, for every fixed $h\ge 3$,
\textsc{$\{P_h\}$-Minor-Free Deletion} parameterized by the feedback vertex set number admits no polynomial kernel
unless $\mathsf{NP}\subseteq\mathsf{coNP}/\mathsf{poly}$.
We adapt their gadget to encode the choices in our reduction from {\permset}.

The key point is to make the choices made by the individual gadgets consistent while keeping the constructed graph close to a forest.
We use only $2k$ shared vertices, one for each row and one for each column, through which the gadgets interact.
Two distinct choices using the same row or the same column are forced to create a $P_h$.
This contrasts with permutation-based lower-bound constructions for treewidth~\cite{BasteST2020Hitting,DBLP:journals/dam/Pilipczuk17}, where the structure enforcing consistency is spread throughout the construction.
In our reduction, all interactions between gadgets go through the same $2k$ shared vertices.
Deleting these $2k$ vertices leaves a disjoint union of trees, so $\fvs(G)\le 2k$, while the row gadgets yield $k$ pairwise vertex-disjoint cycles and hence $\fvs(G)\ge k$.
Thus, $\fvs(G)=\Theta(k)$.
Moreover, every component remaining after deleting the shared
vertices is a tree on $O_h(k)$ vertices and therefore has
treedepth $O_h(\log k)$.
Placing the $2k$ shared vertices above these trees gives
$\td(G)\le 2k+O_h(\log k)$.
Hence, the reduction bounds both parameters simultaneously,
with $\fvs(G)+\td(G)=\Theta_h(k)$,
which yields the lower bound in \Cref{thm:lower}.

The restriction $h \ge 6$ is inevitable for this construction.
Several gadgets may make the same consistent choice $(i,j)$; three such occurrences already give rise to a $K_{2,3}$, which contains a $P_5$.
Appendix~\ref{app:path-five} explains this issue in detail and explicitly gives a \Yes-instance for which it cannot be avoided by choosing different outputs.

\subparagraph*{Algorithm.}
Our algorithm first computes a minimum vertex cover $X$
and guesses which vertices of $X$ belong to a solution.
The vertices outside $X$ form an independent set.
After fixing the guess, we group these vertices by their
neighborhoods in the remaining part of $X$.

Two sparsity properties of fixed-minor-free graphs are crucial here: they have linearly many edges~\cite{Mader1968Homomorphiesatze}, and the number of different ways in which vertices can attach to a fixed vertex set is also linear in the size of that set~\cite{ReidlSVS2019Characterising}.
The edge bound implies that every feasible branch has only
$O_{\mathcal F}(\vc)$ vertices outside $X$ whose neighborhoods
in the remaining part of $X$ have size greater than a suitable
constant depending only on $\mathcal F$.
We keep all these vertices.
Among the other vertices, linear neighborhood complexity bounds
the number of neighborhood types by $O_{\mathcal F}(\vc)$.

For each of these types, we keep only $O_{\mathcal F}(1)$ vertices.
An exchange argument shows that some optimal solution deletes only a constant number of vertices of each type in its branch.
A bounded-degree subgraph containing a forbidden minor uses only a constant number of vertices of each such type, so keeping sufficiently many additional vertices preserves $\mathcal F$-minor-freeness after the deletions.
The resulting graph has $O_{\mathcal F}(\vc)$ vertices.
Enumerating deletion sets and testing minor containment
in this graph gives the claimed running time.

\subparagraph*{Organization.}
In \cref{sec:prelims}, we introduce notation and preliminaries.
\cref{sec:fvs-lower-bound} contains the proof of \cref{thm:lower} and \Cref{sec:vc-algo} contains the proof of \cref{thm:upper}.
We conclude in \cref{sec:conclusion}.

\section{Preliminaries}\label{sec:prelims}

Throughout this paper, $\F$ is a non-empty finite set of finite graphs,
and every graph is simple, finite, and has at least one vertex.
$P_n$ denotes the path on $n$ vertices.
A path of length $k$ contains $k$ edges and $k+1$ vertices.

The \emph{vertex cover number} $\vc(G)$ is the minimum size
of a set $X\subseteq V(G)$ such that $G-X$ is edgeless.
The \emph{feedback vertex set number} $\fvs(G)$ is the minimum size of a set $X\subseteq V(G)$ such that $G-X$ is a forest.
For a graph $G$ and a vertex $v\in V(G)$, we write $N_G(v)$ for the set of
neighbors of $v$ in $G$, and $\Delta(G)$ for the maximum degree of $G$.
For $X\subseteq V(G)$, $G-X$ denotes the graph obtained from $G$ by deleting
the vertices of $X$.

A \emph{minor model} of a graph $H$ in $G$ is a family
$\{B_x\}_{x\in V(H)}$ of pairwise disjoint nonempty sets
$B_x\subseteq V(G)$, called \emph{branch sets}, such that each $G[B_x]$ is
connected and, for every edge $xy\in E(H)$, some edge of $G$ joins $B_x$
and $B_y$. A graph $G$ contains $H$ as a minor if and only if $G$ has a
minor model of $H$.
We write $H\preceq_m G$ if $G$ contains $H$ as a minor.
A graph is \emph{$\F$-minor-free} if it contains no graph
of $\F$ as a minor.
A \emph{$P_h$-deletion set} of $G$ is a set $X\subseteq V(G)$
such that $G-X$ is $P_h$-minor-free.

We recall the definition of treedepth~\cite{NesetrilM2006TreeDepth}.
\begin{definition}
Let $G$ be a graph.
An \emph{elimination forest} of $G$ is a rooted forest $F$ with $V(F)=V(G)$ such that, for every edge $uv\in E(G)$, one of $u$ and $v$ is an ancestor of the other in $F$.
The \emph{height} of $F$ is the maximum number of vertices on a root-to-leaf path in $F$.
The \emph{treedepth} of $G$, denoted by $\td(G)$, is the minimum height of an elimination forest of $G$.
\end{definition}

We also use the following observation.
\begin{lemma}\label{lem:path-minor}
  For every graph $Q$ and integer $h\ge 1$, $P_h\preceq_m Q$ if and only
  if $Q$ has a path on at least $h$ vertices.
\end{lemma}

In \Cref{sec:vc-algo}, we test whether a graph on $O_{\F}(\vc(G))$ vertices contains a graph of $\F$ as a minor.
For this purpose, the following elementary bound is sufficient.
We note that much stronger polynomial-time algorithms are known for testing containment of a fixed minor, following the Graph Minors work of Robertson and Seymour~\cite{RobertsonS1995Graph}.
\begin{lemma}\label{lem:algo-minor-testing}
For every fixed finite family $\F$, there is an algorithm that, given an
$m$-vertex graph $G$, decides in time $2^{O_{\F}(m)}$ whether $G$ contains
a graph of $\F$ as a minor.
\end{lemma}

\begin{proof}
It suffices to test each $H\in\F$ separately, since $\F$ is fixed and finite.
We enumerate all mappings $\phi:V(G)\to V(H)\cup\{\bot\}$. For each
$x\in V(H)$, the set $\phi^{-1}(x)$ is the candidate branch set of $x$, and
vertices mapped to $\bot$ are unused. Since $\phi$ is a function, the
branch sets are pairwise disjoint. For each mapping, we check whether every
$\phi^{-1}(x)$ is nonempty and induces a connected subgraph of $G$, and
whether for every edge $xy\in E(H)$ some edge of $G$ joins
$\phi^{-1}(x)$ and $\phi^{-1}(y)$. If these conditions hold, the sets
$\phi^{-1}(x)$ form a minor model of $H$ in $G$. Conversely, if
$\{B_x\}_{x\in V(H)}$ is a minor model of $H$ in $G$, then the mapping with
$\phi(v)=x$ for $v\in B_x$ and $\phi(v)=\bot$ otherwise satisfies them.

There are $(|V(H)|+1)^m$ mappings, and each can be checked in time
$m^{O(1)}$. Since $|V(H)|$ is bounded by a constant depending only on
$\F$, the total running time is
$|\F|\cdot(\max_{H\in\F}|V(H)|+1)^m\cdot m^{O(1)} = 2^{O_{\F}(m)}$.
\end{proof}

In \Cref{sec:vc-algo}, we use the following sparsity properties of
$\F$-minor-free graphs.

\begin{lemma}\label{lem:prelim-linear-bound}
There are positive integers $c_1$ and $c_2$, depending only on $\F$, such
that the following holds for every $\F$-minor-free graph $P$:
\begin{enumerate}
  \item $|E(P)| \le c_1 |V(P)|$, and
  \item for every nonempty $X \subseteq V(P)$ and every
        $Y \subseteq V(P)\setminus X$, we have
        $|\{N_P(v)\cap X : v\in Y\}| \le c_2 |X|$.
\end{enumerate}
\end{lemma}

\begin{proof}
Fix $H\in\F$ and let $t=|V(H)|$. Every $\F$-minor-free graph is
$H$-minor-free, and hence $K_t$-minor-free, since $H$ is a subgraph of
$K_t$. Item~1 then follows from Mader's
theorem~\cite{Mader1968Homomorphiesatze}, which bounds the number of edges
of $K_t$-minor-free graphs linearly in the number of vertices. For item~2,
the class of $H$-minor-free graphs has bounded
expansion~\cite{DBLP:books/daglib/0030491}, and every class of bounded expansion
has linear neighborhood complexity by the theorem of Reidl, S\'anchez
Villaamil, and Stavropoulos~\cite{ReidlSVS2019Characterising}.
Their bound
counts the traces $N_P(v)\cap X$ over all $v\in V(P)$, so it also applies
to $v\in Y$; the assumption $X\neq\emptyset$ lets the bound absorb the
empty trace.
\end{proof}

\noindent
Since $\F$ is fixed, $c_1$ and $c_2$ are constants, and they are absorbed
into the $O_{\F}(\cdot)$ notation. Our algorithm, however, uses thresholds
derived from these constants, so we need them to be computable. For this reason we give explicit
upper bounds on these constants, which are given in \Cref{app:constants}.

We also use the following result by Fomin, Jansen, and Pilipczuk~\cite{FominJP2014Preprocessing}.
Note that the following statement only states part of the guarantees provided by Proposition~1 of \cite{FominJP2014Preprocessing}.

\begin{lemma}[{\cite[Proposition~1]{FominJP2014Preprocessing}}]
\label{lem:bounded-degree-minor-witness}
Let $G$ and $H$ be graphs.
If $G$ contains $H$ as a minor, then there exists a subgraph $G^* \subseteq G$ that contains $H$ as a minor and satisfies $\Delta(G^*) \leq \Delta(H)$.
\end{lemma}

\section{Superexponential Lower Bound for FVS Parameterization}\label{sec:fvs-lower-bound}

In this section, we prove the following theorem by reducing from {\permset}.

\mainlowerbound*

Throughout this section, fix an integer $h\ge 6$.
\subsection{A Tree Used in the Reduction}
\label{subsec:tree-properties}

We use a tree with designated vertices $y_1,\ldots,y_q$,
which we call its \emph{output vertices}.
Intuitively, the optimal $\{P_h\}$-deletion set
of the tree encodes choosing one out of $q$ choices.
The following lemma states the properties needed for the reduction.
We give its full construction and proof in \Cref{sec:construction-base}.
The construction of the tree follows the same general approach as that of Donkers and Jansen~\cite{DonkersJ2021Turing}.

\begin{lemma}\label{lem:tree-properties}
For every integer $q\ge 1$, we can construct in $O_h(q)$ time
a subcubic tree $T_{h,q}$ with $O_h(q)$ vertices and distinct
output vertices $y_1,\ldots,y_q$ such that:
\begin{enumerate}
  \item Every $P_h$-deletion set of $T_{h,q}$ has size at least $4q-2$.
  \item Every $P_h$-deletion set of size $4q-2$ leaves
    at least one output vertex undeleted.
  \item For every $j\in[q]$, there exists a set
    $S_j\subseteq V(T_{h,q})$ of size $4q-2$ such that
    $T_{h,q}-S_j$ is $P_h$-minor-free,
    $y_j$ is its only output vertex, and $y_j$ is isolated.
\end{enumerate}
\end{lemma}

Thus, deleting exactly $4q-2$ vertices from a copy of $T_{h,q}$ must leave an output vertex.
Conversely, we can choose any output vertex to be the only one
left, with no remaining neighbor inside the tree.
In the reduction, edges between a copy of $T_{h,q}$ and the rest of the graph will be incident only to its output vertices.

\subsection{The Construction}
\label{subsec:reduction}

Let $\Gamma$ be an instance of {\permset} with parameter $k\ge 2$.
We may delete every edge whose endpoints lie in the same row or the same column, since a permutation independent set cannot contain both endpoints of such an edge.
Hence, assume that every edge $\{(i,j),(i',j')\}\in E(\Gamma)$ satisfies $i\ne i'$ and $j\ne j'$.
Let $m:=|E(\Gamma)|$; then $m=O(k^4)$.

For each row $i$, we use a copy of $T_{h,k}$ whose output vertices
correspond to the choices $(i,1),\ldots,(i,k)$.
For each edge $\{(i,j),(i',j')\}$, we use another tree whose outputs
correspond to the choices in row $i$ other than $(i,j)$ and
the choices in row $i'$ other than $(i',j')$.
Requiring an output to remain in this tree will ensure that
the two endpoints of the edge are not both selected.

Let $L:=\lfloor(h-4)/2\rfloor$.
Since $h\ge 6$, we have $L\ge 1$ and $2L+3<h\le 2L+5$.
We construct $G$ as follows.
All vertices introduced below are distinct unless stated otherwise.

\begin{enumerate}
  \item Create the sets
    $R:=\{r_1,\ldots,r_k\}$ and $C:=\{c_1,\ldots,c_k\}$.
    The vertices $r_i$ and $c_j$ correspond to row $i$
    and column $j$, respectively.

  \item Attach a pendant path of length $L$ to each vertex
    of $R\cup C$, using new vertices for the rest of each path.

  \item For each $i\in[k]$, create a copy $T_i^{\mathrm{row}}$
    of $T_{h,k}$.
    Denote its output vertices by $y_{i,1},\ldots,y_{i,k}$.
    Give $y_{i,j}$ the label $(i,j)$ and add the edges
    $r_i y_{i,j}$ and $y_{i,j}c_j$ for every $j\in[k]$.

  \item For each edge $e=\{(i,j),(i',j')\}\in E(\Gamma)$,
    create a copy $T^e$ of $T_{h,2k-2}$.
    Denote its output vertices by
    $x_{i,a}^e$ for $a\in[k]\setminus\{j\}$ and
    $x_{i',a}^e$ for $a\in[k]\setminus\{j'\}$,
    assigning these names to the outputs in any fixed order.
    Give each $x_{s,t}^e$ the label $(s,t)$ and add the edges
    $r_s x_{s,t}^e$ and $x_{s,t}^e c_t$.
\end{enumerate}

We call $T_i^{\mathrm{row}}$ a \emph{row gadget}
and $T^e$ an \emph{edge gadget}.
These gadgets are pairwise vertex-disjoint and pairwise non-adjacent.
Only their output vertices have neighbors outside the gadget,
and an output vertex with label $(i,j)$ has exactly $r_i$ and $c_j$
as its neighbors outside the gadget.

\begin{figure}[htb]
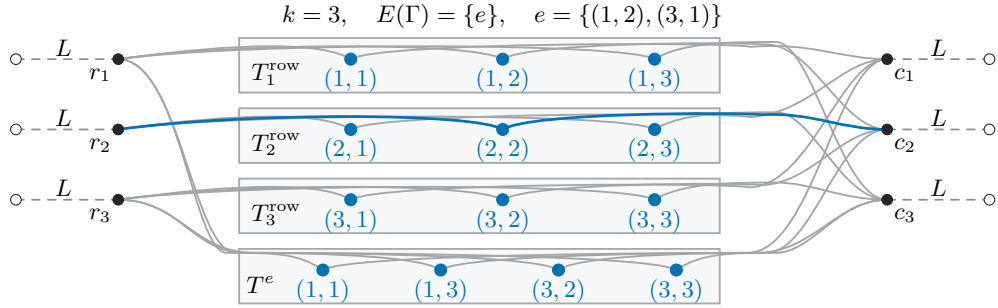

  \centering
  \GlobalGraphPaper
  \caption{The construction for $k=3$ and
    $E(\Gamma)=\{\{(1,2),(3,1)\}\}$.
    Each box contains a copy of $T_{h,q}$;
    only its output vertices are drawn.
    An output labeled $(i,j)$ is adjacent to $r_i$ and $c_j$.
    All outputs and edges between the gadgets and $R\cup C$
    are shown.
    The two edges incident to the output labeled $(2,2)$
    are highlighted.
    Dashed segments represent pendant paths of length $L$.}
  \label{fig:global-construction}
\end{figure}

Set $B:=k(4k-2)+m(8k-10)$.
By \Cref{lem:tree-properties}, every $P_h$-deletion set of $G$
contains at least $4k-2$ vertices from each row gadget
and at least $8k-10$ vertices from each edge gadget.
Since the gadgets are vertex-disjoint, every such deletion set
has size at least $B$.
The reduction produces the instance $(G,B)$.

\subsection{Correctness}
\label{subsec:reduction-correctness}

Throughout this subsection, let $(G,B)$ be the instance constructed
in \Cref{subsec:reduction} from an instance $\Gamma$ of {\permset}
with vertex set $[k]\times[k]$.
We prove the following equivalence.

\begin{theorem}\label{thm:reduction}
The graph $\Gamma$ has a permutation independent set if and only if
$G$ has a $P_h$-deletion set of size at most $B$.
\end{theorem}

\begin{figure}[htb]
  \centering
  \FVSProofMap
  \caption{The two directions of the proof of \Cref{thm:reduction}.}
  \label{fig:fvs-proof-map}
\end{figure}

We first observe how a deletion set of size at most $B$ intersects the gadgets.

\begin{lemma}\label{lem:deletions-per-gadget}
Let $X\subseteq V(G)$ with $|X|\le B$, and suppose that $G-X$ is $P_h$-minor-free.
Then:
\begin{enumerate}
  \item $|X\cap V(T_i^{\mathrm{row}})|=4k-2$
    for every $i\in[k]$;
  \item $|X\cap V(T^e)|=8k-10$
    for every $e\in E(\Gamma)$;
  \item no vertex of $R\cup C$ or of their attached pendant paths belongs to $X$; and
  \item every row gadget and every edge gadget has
    an output vertex outside $X$.
\end{enumerate}
\end{lemma}

\begin{proof}
The intersection of $X$ with each gadget is a $P_h$-deletion set
of that gadget.
By \Cref{lem:tree-properties}, these intersections have size
at least $4k-2$ for row gadgets and $8k-10$ for edge gadgets.
Their sum is at least $B$.
Since $|X|\le B$ and the gadgets are vertex-disjoint,
all these bounds hold with equality, and $X$ contains no vertex
outside the gadgets.
This proves the first three statements.

Each gadget is therefore a copy of $T_{h,q}$
from which exactly $4q-2$ vertices are deleted.
The last statement follows from \Cref{lem:tree-properties}.
\end{proof}

\subsubsection{Soundness}

\begin{proposition}\label{prop:soundness}
If $G$ has a $P_h$-deletion set of size at most $B$,
then $\Gamma$ has a permutation independent set.
\end{proposition}

\begin{figure}[tbp]
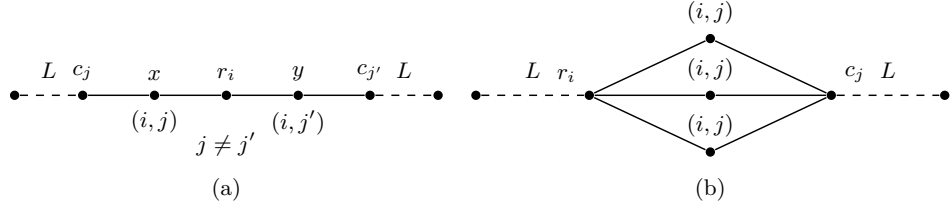

  \centering
  \CorrectnessPictures
  \caption{(a) Outputs with labels $(i,j)$ and $(i,j')$,
    where $j\ne j'$, give a path on $2L+5\ge h$ vertices.
    (b) Three outputs with the same label, together with
    their two neighbors and the attached paths.
    The longest path has $2L+3<h$ vertices.
    Dashed segments represent paths of length $L$.}
  \label{fig:conflict}
  \label{fig:repeated-label}
\end{figure}

\begin{proof}
Let $X$ be a $P_h$-deletion set of $G$ with $|X|\le B$.

Suppose that two output vertices $x,y\notin X$ have labels
$(i,j)$ and $(i,j')$, where $j\ne j'$.
By \Cref{lem:deletions-per-gadget}, the vertices
$c_j,r_i,c_{j'}$ and their attached paths remain in $G-X$.
Hence, $c_j-x-r_i-y-c_{j'}$ is a path in $G-X$.
Appending the pendant paths at $c_j$ and $c_{j'}$
gives a path on $2L+5\ge h$ vertices,
contradicting \Cref{lem:path-minor}.
Similarly, two outputs outside $X$ cannot have labels
$(i,j)$ and $(i',j)$ with $i\ne i'$.
See \Cref{fig:conflict}(a).

For every $i\in[k]$, the row gadget $T_i^{\mathrm{row}}$
has an output outside $X$.
Choose one such output and write its label as $(i,\pi(i))$.
By the preceding argument, every output outside $X$
whose label has first coordinate $i$ has second coordinate
$\pi(i)$.
Moreover, $\pi(i)\ne\pi(i')$ whenever $i\ne i'$.
Thus, $\pi:[k]\to[k]$ is a permutation.

It remains to show that $\{(i,\pi(i)):i\in[k]\}$ is independent.
Fix an edge $e=\{(i,j),(i',j')\}$ of $\Gamma$.
By \Cref{lem:deletions-per-gadget}, some output of $T^e$
lies outside $X$.
Its label is either $(i,a)$ with $a\ne j$ or
$(i',a)$ with $a\ne j'$.
In the first case, $a=\pi(i)$, so $\pi(i)\ne j$.
In the second case, $a=\pi(i')$, so $\pi(i')\ne j'$.
Therefore, the permutation does not select both endpoints of $e$.
This holds for every edge of $\Gamma$, proving the claim.
\end{proof}

\subsubsection{Completeness}

\begin{proposition}\label{prop:completeness}
If $\Gamma$ has a permutation independent set,
then $G$ has a $P_h$-deletion set of size $B$.
\end{proposition}

\begin{proof}
Let $\{(i,\pi(i)):i\in[k]\}$ be a permutation independent set
of $\Gamma$.

For each row gadget $T_i^{\mathrm{row}}$, use
\Cref{lem:tree-properties} to choose a set $X_i^{\mathrm{row}}$
of size $4k-2$ such that
$T_i^{\mathrm{row}}-X_i^{\mathrm{row}}$ is $P_h$-minor-free
and $y_{i,\pi(i)}$ is its only output vertex and is isolated.

Now fix an edge $e=\{(i,j),(i',j')\}$ of $\Gamma$.
Since the permutation independent set does not contain
both endpoints of $e$, we have $\pi(i)\ne j$ or $\pi(i')\ne j'$.
Thus, $T^e$ has an output with label $(s,\pi(s))$
for some $s\in\{i,i'\}$.
Choose one such output.
By \Cref{lem:tree-properties}, there is a set $X^e$
of size $8k-10$ such that $T^e-X^e$ is $P_h$-minor-free
and the chosen output is its only output vertex and is isolated.

Let $X$ be the union of all sets $X_i^{\mathrm{row}}$ and $X^e$.
The gadgets are pairwise vertex-disjoint, so $|X|=B$.
We show that $G-X$ is $P_h$-minor-free.

Within each gadget, its only remaining output is isolated.
Since only output vertices have neighbors outside the gadget,
every component not containing that output is also a component
of $G-X$ and is $P_h$-minor-free.

Every remaining output vertex has label $(i,\pi(i))$ for some $i\in[k]$.
Fix $i$, and let $s_i\ge 1$ be the number of outputs with this label.
Together with $r_i$ and $c_{\pi(i)}$, these outputs induce
a copy of $K_{2,s_i}$.
Including the pendant paths attached to $r_i$ and $c_{\pi(i)}$
gives a component of $G-X$; see \Cref{fig:repeated-label}(b).
Since $\pi$ is a permutation, these components are disjoint
for different values of $i$.

Consider a path in one of these components.
If it uses vertices from both pendant paths outside $R\cup C$,
it contains at most $2L+3$ vertices.
If it uses vertices outside $R\cup C$ from exactly one
of the pendant paths, it contains at most $L+4$ vertices.
Otherwise, it is contained in $K_{2,s_i}$
and contains at most five vertices.
Since $L\ge 1$, in all three cases, there are at most $2L+3<h$ vertices.
Hence, these components are also $P_h$-minor-free
by \Cref{lem:path-minor}.

We have checked for every component of $G-X$.
Therefore, $G-X$ is $P_h$-minor-free.
\end{proof}

\begin{proof}[Proof of \Cref{thm:reduction}]
The two directions follow from
\Cref{prop:soundness,prop:completeness}.
\end{proof}

\subsection{Size and Parameter Bounds of the Constructed Graph}
\label{subsec:hardness}

\begin{proposition}\label{prop:size-fvs}
The constructed graph $G$ has $N=O_h(k^5)$ vertices and satisfies
$k\le\fvs(G)\le 2k$ and $\td(G)\le 2k+O_h(\log k)$.
Moreover, $R\cup C$ is a feedback vertex set of size $2k$.
\end{proposition}

\begin{proof}
There are $k$ row gadgets and $m=O(k^4)$ edge gadgets,
each with $O_h(k)$ vertices.
The vertices of $R\cup C$ and their attached pendant paths contribute
$2k(L+1)=O_h(k)$ vertices.
Thus, $N=O_h(k^2+mk)=O_h(k^5)$.

Deleting $R\cup C$ leaves the row gadgets, the edge gadgets,
and the attached paths with their endpoints in $R\cup C$ removed.
These are all trees, so $R\cup C$ is a feedback vertex set
and $\fvs(G)\le 2k$.

For the lower bound, consider the output vertices $y_{i,1}$ and $y_{i,2}$
in each row gadget $T_i^{\mathrm{row}}$.
They exist because $k\ge 2$.
The unique path between them in $T_i^{\mathrm{row}}$,
together with the edges $r_i y_{i,1}$ and $r_i y_{i,2}$,
forms a cycle.
These cycles are vertex-disjoint for different rows.
Every feedback vertex set must meet each cycle,
so $\fvs(G)\ge k$.

Finally, every component of $G-(R\cup C)$ is a tree
on $O_h(k)$ vertices.
Recursively choosing a centroid gives each such tree
an elimination forest of height $O_h(\log k)$.
Place the $2k$ vertices of $R\cup C$ on a single rooted path
and attach the roots of these forests as children of its last vertex.
This gives an elimination forest of $G$ of height
at most $2k+O_h(\log k)$.
\end{proof}

\begin{proof}[Proof of \Cref{thm:lower}]
Suppose that there is an algorithm for
\textsc{$\{P_h\}$-Minor-Free Deletion} running in time
$2^{o(t\log t)}N^{O(1)}$, where
$t=\fvs(G)+\td(G)$,
when a minimum feedback vertex set is given as a part of the input.

Given an instance $(\Gamma,k)$ of {\permset},
construct $(G,B)$ as above.
The construction takes polynomial time for fixed $h$.
By \Cref{prop:size-fvs}, we have
$N=O_h(k^5)$ and $t=\Theta_h(k)$.

The set $R\cup C$ is a feedback vertex set of size $2k$.
A minimum feedback vertex set of $G$ can therefore be computed
in $2^{O(k)}N^{O(1)}$ time~\cite{KociumakaP2014Faster}.
Give this set to the assumed algorithm as a part of the input and run it on $(G,B)$.
By \Cref{thm:reduction}, its answer determines whether
$\Gamma$ has a permutation independent set.

Since $t=\Theta_h(k)$ and $N=O_h(k^5)$,
the assumed algorithm takes
$2^{o(k\log k)}k^{O_h(1)}$ time.
Computing the minimum feedback vertex set takes
$2^{O(k)}k^{O_h(1)}$ time,
so the total running time is
$2^{o(k\log k)}k^{O_h(1)}$.
This contradicts \Cref{lem:permutation-eth} under ETH.
\end{proof}

\subsection{Construction of $T_{h,q}$}
\label{sec:construction-base}

We now construct the tree in \Cref{lem:tree-properties}.
The construction follows that of Donkers and Jansen~\cite{DonkersJ2021Turing}, but we fully describe it here for the completeness of the paper.

Let $d:=h-2$.
Start with the path $v-b-a-w$.
Attach a path of length $d$ to $a$, with other endpoint $u$,
and another path of length $d$ to $b$, with other endpoint $z$.
The two attached paths use new vertices and are otherwise disjoint.
Let $T$ be the resulting tree.
It has $4+2d=2h$ vertices and maximum degree three.

\begin{figure}[htbp]
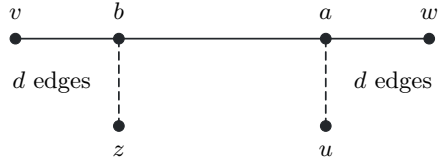

  \centering
  \BaseTreePicture
  \caption{The tree $T$.
    Each dashed segment represents a path of $d=h-2$ edges.}
  \label{fig:base}
\end{figure}

For $q\ge 1$, take copies $M_1,\ldots,M_q$ of $T$.
Use subscripts to denote their vertices:
$a_i,b_i,u_i,z_i,w_i,v_i\in V(M_i)$.
Also take copies $N_1,\ldots,N_{q-1}$ of $T$,
with corresponding vertices
$a_i',b_i',u_i',z_i',w_i',v_i'$.
For every $i\in[q-1]$, identify $v_i'$ with $w_i$
and $w_i'$ with $u_{i+1}$.

\begin{figure}[htbp]
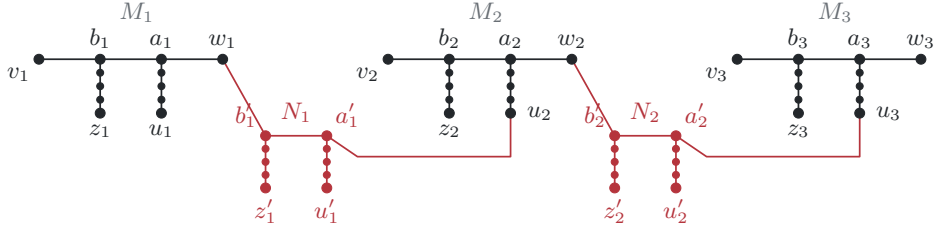

  \centering
  \ChainPaper
  \caption{For $h=6$ and $q=3$, the copies $M_1,M_2,M_3$
    are black and $N_1,N_2$ are red.
    The identified vertices satisfy
    $v_i'=w_i$ and $w_i'=u_{i+1}$.
    All subdivision vertices are shown.}
  \label{fig:chain}
\end{figure}

For each $i\in[q]$, add a new vertex $y_i$ adjacent to $v_i$.
Also attach a path of length $d$ to $v_i$,
using new vertices and denoting its other endpoint by $e_i$.
The resulting graph is $T_{h,q}$,
and its output vertices are $y_1,\ldots,y_q$.

\begin{figure}[htbp]
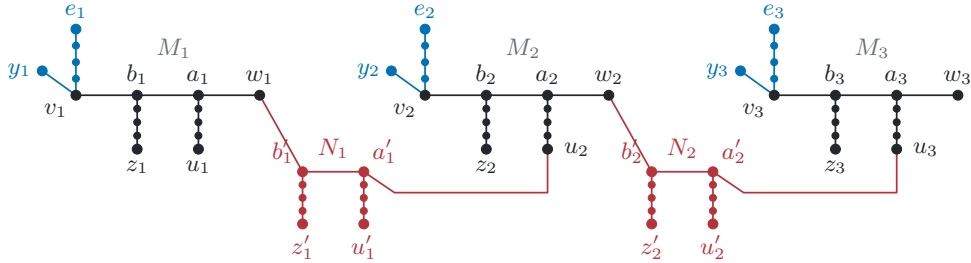

  \centering
  \SelectorPaper
  \caption{The tree $T_{6,3}$.
    The paths ending at $e_1,e_2,e_3$
    and the edges to the output vertices are blue.}
  \label{fig:gadget}
\end{figure}

\begin{proof}[Proof of \Cref{lem:tree-properties}]
The copies of $T$ are joined in the order
$M_1,N_1,M_2,\ldots,N_{q-1},M_q$,
with consecutive copies sharing one vertex.
Thus, their union is a tree.
Adding the output vertices and the paths ending at $e_i$
preserves this property.
Each identification joins two leaves,
and adding $y_i$ and the path ending at $e_i$
increases the degree of $v_i$ from one to three.
Hence, $T_{h,q}$ is subcubic.

The $2q-1$ copies of $T$ contribute $2h(2q-1)$ vertices.
The identifications reduce this number by $2(q-1)$.
For each $i$, the path ending at $e_i$ contributes $d$ new vertices,
and $y_i$ contributes one.
Therefore,
$|V(T_{h,q})|
=2h(2q-1)-2(q-1)+q(d+1)
=(5h-3)q-2h+2=O_h(q)$.
The construction takes $O_h(q)$ time.

To prove the first property, we find $4q-2$ vertex-disjoint copies of $P_h$ in $T_{h, q}$.
For each $i\in[q]$, take the path $a_i-b_i\cdots z_i$.
For each $i\in[q-1]$, take the two paths
$u_i'\cdots a_i'-w_i'$ and $v_i'-b_i'\cdots z_i'$.
These are $3q-2$ pairwise vertex-disjoint paths,
each on $h$ vertices.
None contains a vertex $v_i$.
We may therefore also take $e_i\cdots v_i-y_i$
for each $i\in[q]$.
This gives $4q-2$ vertex-disjoint copies of $P_h$,
so every $P_h$-deletion set has size at least $4q-2$.
See \Cref{fig:packing-with-outputs}.

\begin{figure}[htbp]
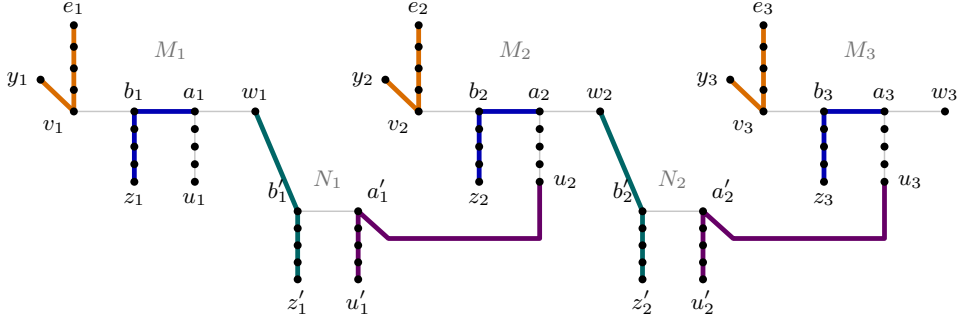

  \centering
  \PackingWithOutputsPicture
  \caption{Ten vertex-disjoint copies of $P_6$ in $T_{6,3}$.
    The highlighted paths are
    $a_i-b_i\cdots z_i$ and $e_i\cdots v_i-y_i$
    for $i\in[3]$, and
    $v_i'-b_i'\cdots z_i'$ and $u_i'\cdots a_i'-w_i'$
    for $i\in[2]$.}
  \label{fig:packing-with-outputs}
\end{figure}

For the second property, remove all output vertices.
The remaining graph still contains $3q-1$
vertex-disjoint copies of $P_h$:
take $u_i\cdots a_i-w_i$ and $z_i\cdots b_i-v_i$
for every $i\in[q]$, and take
$z_i'\cdots b_i'-a_i'$ for every $i\in[q-1]$.
Thus, a $P_h$-deletion set containing all $q$ outputs
must have size at least $q+(3q-1)=4q-1$.
In particular, a deletion set of size $4q-2$
cannot contain all outputs.
See \Cref{fig:packing-without-outputs}.

\begin{figure}[htbp]
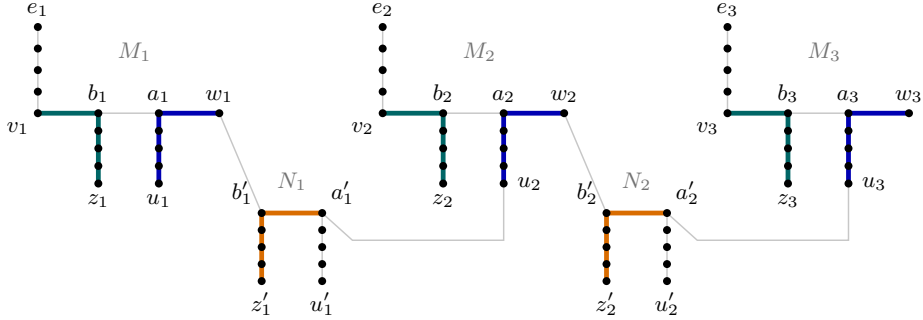

  \centering
  \PackingWithoutOutputsPicture
  \caption{Eight vertex-disjoint copies of $P_6$ in
    $T_{6,3}-\{y_1,y_2,y_3\}$.
    The highlighted paths are
    $u_i\cdots a_i-w_i$ and $z_i\cdots b_i-v_i$
    for $i\in[3]$, and
    $z_i'\cdots b_i'-a_i'$ for $i\in[2]$.}
  \label{fig:packing-without-outputs}
\end{figure}

For the third property, fix $j\in[q]$.
Let $S_j$ consist of the following vertices:
\begin{itemize}
  \item $y_i$ for every $i\ne j$;
  \item $b_i,w_i,a_i'$ for every $i<j$;
  \item $a_j,v_j$; and
  \item $b_i,u_i,b_{i-1}'$ for every $i>j$.
\end{itemize}
These vertices are distinct, so
$|S_j|=(q-1)+3(j-1)+2+3(q-j)=4q-2$.
The only output outside $S_j$ is $y_j$,
and it is isolated because $v_j\in S_j$.
See \Cref{fig:chosen-output}.

\begin{figure}[htbp]
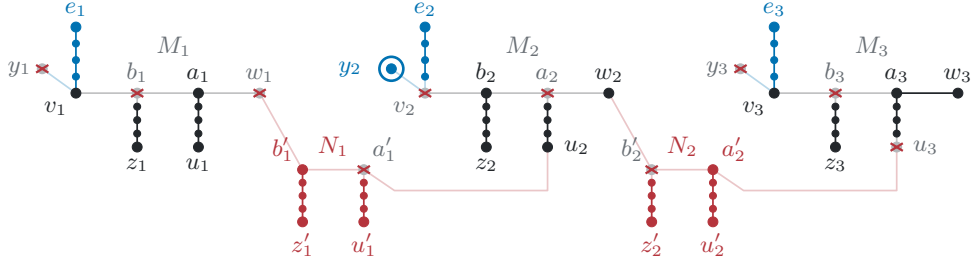

  \centering
  \CanonicalPaper
  \caption{The ten vertices of $S_2$ in $T_{6,3}$
    are marked by $\times$.
    Edges incident to deleted vertices are faded.
    The only remaining output vertex is $y_2$, which is an isolated vertex.}
  \label{fig:chosen-output}
\end{figure}

We verify that every component of $T_{h,q}-S_j$
is a path on at most $h-1$ vertices.

Within $M_i$, we delete $\{b_i,w_i\}$ if $i<j$,
$\{a_i,v_i\}$ if $i=j$, and $\{b_i,u_i\}$ if $i>j$.
Within $N_i$, we delete $\{a_i',v_i'\}$ if $i<j$
and $\{b_i',w_i'\}$ if $i\ge j$,
using the identifications $v_i'=w_i$ and $w_i'=u_{i+1}$.
Deleting any of $\{b,w\}$, $\{a,v\}$, or $\{b,u\}$
from $T$ leaves only path components on at most $d+1=h-1$ vertices.

These paths do not join into longer paths through the identified
vertices.
At $w_i=v_i'$, the identified vertex is deleted if $i<j$.
If $i\ge j$, then $b_i'$ is deleted,
so the identified vertex has no remaining neighbor in $N_i$.
Similarly, at $w_i'=u_{i+1}$, the identified vertex is deleted
if $i\ge j$.
If $i<j$, then $a_i'$ is deleted,
so it has no remaining neighbor in $N_i$.

It remains to consider the paths ending at $e_i$.
If $i\ne j$, both $b_i$ and $y_i$ are deleted.
Thus, the path from $v_i$ to $e_i$ is a component
on $d+1=h-1$ vertices.
If $i=j$, the vertex $v_j$ is deleted,
leaving a path on $d=h-2$ vertices and the isolated vertex $y_j$.
Consequently, every component of $T_{h,q}-S_j$
is a path on at most $h-1$ vertices.
By \Cref{lem:path-minor}, the graph is $P_h$-minor-free.
\end{proof}

\section{Algorithm Parameterized by Vertex Cover Number}\label{sec:vc-algo}

\subsection{Overview and Setup}\label{subsec:vc-setup}
If $\F$ contains an edgeless graph, let $r$ be the minimum
number of vertices in such a graph.
Then every $\F$-minor-free graph has fewer than $r$ vertices.
Let $\F':=\{H\in\F:E(H)\neq\emptyset\}$.
Then $(G,k)$ is a \Yes-instance for $\F$ if and only if
$k\ge |V(G)|-r+1$ and $(G,k)$ is a \Yes-instance for $\F'$.
Hence, throughout the remainder of this section, we assume that every graph in $\F$ contains at least one edge.

Let $(G,k)$ be an instance of {\problemname}.
Compute a minimum vertex cover $X$ of $G$ in time
$2^{O(\vc(G))}\cdot |V(G)|^{O(1)}$~\cite{DBLP:conf/stacs/0001N24}, and put
$I:=V(G)\setminus X$ and $t:=|X|=\vc(G)$.
Then $I$ is an independent set.
If $k\ge t$, deleting $X$ gives a solution, since every graph in $\F$ contains an edge.
Hence, assume that $k<t$.

We enumerate all sets $D\subseteq X$ with $|D|\le k$.
For each $D$, let $Y:=X\setminus D$ and $b:=k-|D|$.
We ask whether there exists a set $S\subseteq I$ with $|S|\le b$ such that $G-(D\cup S)$ is $\F$-minor-free.
We call the branch \emph{feasible} if such a set exists.
Since $k<t$, we have $Y\ne\emptyset$.
We discard the branch if $G[Y]$ is not $\F$-minor-free,
since no vertex of $Y$ may be deleted.

We will reduce this question to a graph on $O_{\F}(t)$ vertices.
To do so, we group vertices of $I$ by their neighborhoods in $Y$.

Let $c_1,c_2$ be the constants from \Cref{lem:prelim-linear-bound}, and put $h := \max_{H \in \F} |V(H)|$ and $\Delta_\F := \max_{H \in \F} \Delta(H)$.
For each $v\in I$, define $\Type(v):=N_G(v)\cap Y$.
Let $I_1:=\{v\in I:|\Type(v)|\ge c_1+1\}$ and $I_2:=I\setminus I_1$.
We will include every vertex of $I_1$ in the smaller graph.
For the vertices in $I_2$, let $\Types:=\{\Type(v):v\in I_2\}$, and for $Q \in \Types$, let $B_Q:=\{v\in I_2:\Type(v)=Q\}$.
We call $Q$ a \emph{neighborhood type} and $B_Q$ its
\emph{bucket}.
If $\emptyset\notin\Types$, set $B_\emptyset:=\emptyset$.

\begin{figure}[htbp]
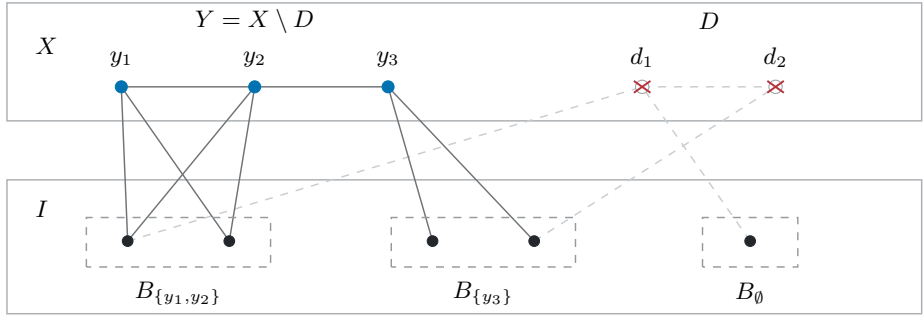

\centering
\BVCBranchFigure{1}
\caption{Neighborhood types in a fixed branch $D\subseteq X$.
Crossed vertices belong to $D$; pale dashed edges disappear in $G-D$.
Each bucket contains vertices with the same neighborhood in $Y=X\setminus D$,
even if their neighborhoods in $D$ differ.}
\label{fig:vc-algo}
\end{figure}

The vertices of each $B_Q$ are false twins in $G-D$: they are pairwise nonadjacent and have the same neighborhood $Q$.
Thus, the graph obtained by deleting vertices from $B_Q$ depends, up to isomorphism, only on how many vertices are deleted.
We will show that keeping a bounded number of vertices from each $B_Q$ is sufficient.
\Cref{fig:vc-algo} illustrates the buckets $B_Q$.

We first show that every feasible branch has $|I_1|=O_{\F}(t)$ and $|\Types|=O_{\F}(t)$.
We then choose an optimal solution and show that, in its branch,
it deletes at most $|Q|-1$ vertices from each bucket $B_Q$
with $Q\ne\emptyset$, and none from $B_\emptyset$.
Finally, we show that keeping $|Q|\Delta_{\F}+h$ undeleted vertices from each bucket is sufficient to preserve $\F$-minor-freeness.
\Cref{fig:vc-proof-map} summarizes the proof.

\begin{figure}[htbp]
  \centering
  \VCProofMap
  \caption{Proof of \Cref{thm:main-vc}. The full description of the algorithm is given in \Cref{alg:f-minor-deletion-vc}.}
  \label{fig:vc-proof-map}
\end{figure}

\subsection{Bounding the Search Space}
\label{subsec:vc-bounds}

We begin with two necessary conditions for a branch to be feasible.

\begin{lemma}\label{lem:vc-high-deg-bound}
If the branch $D$ is feasible, then $|I_1|\le c_1|Y|+b$.
\end{lemma}

\begin{proof}
Let $S\subseteq I$ with $|S|\le b$ and suppose that
$G-(D\cup S)$ is $\F$-minor-free.
Put $J:=I_1\setminus S$, and let $P$ be the bipartite graph on $Y\cup J$ with all edges of $G$ crossing between $Y$ and $J$.
Since $P$ is a subgraph of $G-(D\cup S)$, the graph $P$ is
$\F$-minor-free.

Every vertex of $J$ has at least $c_1+1$ neighbors in $Y$.
By \Cref{lem:prelim-linear-bound}, $(c_1+1)|J| \le |E(P)| \le c_1(|Y|+|J|)$.
Hence $|J|\le c_1|Y|$.
Since at most $b$ vertices of $I_1$ belong to $S$, the claimed bound follows.
\end{proof}

\begin{lemma}\label{lem:vc-type-bound}
If the branch $D$ is feasible, then $|\Types|\le c_2|Y|+b$.
\end{lemma}

\begin{proof}
Let $S\subseteq I$ with $|S|\le b$ and suppose that $G-(D\cup S)$ is $\F$-minor-free.
A type $Q\in\Types$ either has a vertex in $B_Q\setminus S$ or satisfies $B_Q\subseteq S$.

Since $Y\ne\emptyset$, \Cref{lem:prelim-linear-bound} bounds
the number of types $Q$ with $B_Q \setminus S \neq \emptyset$ by $c_2|Y|$.
Since the buckets are pairwise disjoint and nonempty, at most $|S|\le b$ of them are entirely contained in $S$.
Thus $|\Types|\le c_2|Y|+b$.
\end{proof}

By \Cref{lem:vc-high-deg-bound,lem:vc-type-bound}, we may discard a branch if $|I_1|>c_1|Y|+b$ or $|\Types|>c_2|Y|+b$.
In every remaining branch, $|I_1|$ and $|\Types|$ are
$O_{\F}(t)$, since $|Y|\le t$ and $b<t$.
Individual buckets may still be arbitrarily large.
The next lemma shows that some optimal solution deletes only
a bounded number of vertices from each bucket in its branch.

\begin{lemma}\label{lem:vc-representatives}
Suppose that $(G,k)$ is a \Yes-instance and $k<|X|$.
Among all minimum-size solutions, choose $S^*$ maximizing
$|S^*\cap X|$.
Let $D:=S^*\cap X$ and $Y:=X\setminus D$, and define
$\Types$ and $B_Q$ with respect to this $Y$.
Then $|S^*\cap B_Q|\le |Q|-1$ for every nonempty $Q\in\Types$,
and $S^*\cap B_\emptyset=\emptyset$.
\end{lemma}

\begin{proof}
As $|S^*| \le k < |X|$, $Y$ is nonempty.

First, suppose to the contrary that $|Q| \ge 1$ and $|S^* \cap B_Q| \ge |Q|$.
Choose exactly $|Q|$ vertices $Z \subseteq S^* \cap B_Q$ and define $S' := (S^* \setminus Z) \cup Q$.
As $Q \subseteq Y$, $Q \cap S^* = \emptyset$.
So $|S'| = |S^*|$ while $|S' \cap X|  = |S^* \cap X| + |Q|$.
Now, we show that $S'$ is also a solution.
In $G - S'$, vertices of $Z$ are isolated, because their neighbors in $G$ lie in $D \cup Q$, and all of these are deleted by $S'$.
Hence, $G-S'$ is exactly the graph that can be obtained from $G-S^*$ by removing all edges incident to $Q$ and relabeling vertices of $Q$ bijectively by the vertices of $Z$.
Thus, $G-S'$ is isomorphic to a spanning subgraph of $G-S^*$, which implies that $G-S'$ is also $\F$-minor-free.
This contradicts the choice of $S^*$.
Therefore, $|S^* \cap B_Q| \le |Q|-1$.

Next, suppose that there exists a vertex $v \in S^* \cap B_\emptyset$.
Choose a vertex $y \in Y$ and let $S' := (S^* \setminus \{v\}) \cup \{y\}$.
In $G - S'$, $v$ is isolated because $N_G(v) \subseteq D$.
Hence, $G-S'$ is exactly the graph that can be obtained from $G-S^*$ by removing all edges incident to $y$ and relabeling it to $v$.
Thus, $G-S'$ is also $\F$-minor-free, once again a contradiction to the choice of $S^*$.
\end{proof}

Since $|Q|\le c_1$ for every $Q\in\Types$, the solution chosen in \Cref{lem:vc-representatives} deletes
only $O_{\F}(1)$ vertices from each bucket.
We next show that, after these deletions, keeping
$|Q|\Delta_{\F}+h = O_\F(1)$ vertices from a bucket is sufficient
to preserve $\F$-minor-freeness.

\begin{lemma}
\label{lem:vc-false-twin-reduce}
Let $P$ be a graph, and let $T\subseteq V(P)$ be a set of pairwise false twins with common neighborhood $Q$.
Let $T'\subseteq T$ with $|T'|\geq |Q|\Delta_{\F}+h$.
Then $P$ is $\F$-minor-free if and only if $P-(T\setminus T')$ is $\F$-minor-free.
\end{lemma}

\begin{proof}
Since $P - (T \setminus T')$ is a subgraph of $P$, one direction is immediate.

For the converse, we show the contrapositive.
Suppose that $P$ contains a graph $H \in \F$ as a minor.
By \Cref{lem:bounded-degree-minor-witness}, there exists a subgraph $P^* \subseteq P$ such that it contains $H$ as a minor and $\Delta(P^*) \le \Delta(H)$.
Fix a minor model of $H$ in $P^*$ and remove every isolated vertex not used by this model.
Each remaining isolated vertex forms a singleton branch set, so $P^*$ has at most $|V(H)|$ isolated vertices.

Every non-isolated vertex of $T\cap V(P^*)$ has a neighbor in $Q\cap V(P^*)$.
Since $\Delta(P^*)\le\Delta(H)$, there are at most $|Q|\Delta(H)$ such vertices.
Including isolated vertices, we have $|T\cap V(P^*)|\le |Q|\Delta(H)+|V(H)| \le |Q|\Delta_{\F}+h$.
Since $|T'|\geq |Q|\Delta_{\F}+h$, we can replace all vertices of $T\cap V(P^*)$ by vertices of $T'$ injectively.
As the vertices of $T$ are pairwise false twins, they are independent and have the same neighborhood $Q$.
Therefore, this replacement yields a subgraph of $P-(T\setminus T')$ that is isomorphic to $P^*$, which implies that $P-(T\setminus T')$ contains $H$ as a minor.
\end{proof}

\subsection{The Algorithm}

Consider a branch that has not been discarded.
For each nonempty $Q\in\Types$, we keep at most $(|Q|-1)+(|Q|\Delta_{\F}+h)$ vertices from $B_Q$.
If $B_Q$ contains more vertices, then at least $|Q|\Delta_{\F}+h$ of the kept vertices remain after at most $|Q|-1$ deletions.
For $Q=\emptyset$, we keep at most $h$ vertices and allow no deletions.
Thus, for each $Q\in\Types$, choose $R_Q\subseteq B_Q$ with
\begin{equation}\label{eq:vc-representatives}
|R_Q|=
\begin{cases}
\min\{|B_Q|,h\},
    & Q=\emptyset,\\
\min\{|B_Q|,\,
      \underbrace{|Q|-1}_{\text{deletions}}
      +\underbrace{|Q|\Delta_{\F}+h}_{\text{survivors}}\},
    & Q\ne\emptyset.
\end{cases}
\end{equation}

If $\emptyset\notin\Types$, set $R_\emptyset:=\emptyset$.
Let $R:=\bigcup_{Q\in\Types}R_Q$ and $G':=G[Y\cup I_1\cup R]$.

We enumerate sets of vertices to delete from $I_1\cup R$,
while keeping every vertex of $Y$.
Vertices of $I_2\setminus R$ are omitted from $G'$;
they are not included in the deletion set.
Specifically, enumerate $A\subseteq I_1$ and $Z\subseteq R$ subject to
\begin{equation}\label{eq:vc-admissible}
\begin{aligned}
|A|+|Z|&\le b,\\
Z\cap R_\emptyset&=\emptyset,\\
|Z\cap R_Q|&\le |Q|-1
    &&(Q\in\Types\setminus\{\emptyset\}).
\end{aligned}
\end{equation}
We accept if $G'-(A\cup Z)$ is $\F$-minor-free.
In the accepted branch, the corresponding deletion set in the original graph is $D\cup A\cup Z$.
The pseudocode is given in \Cref{alg:f-minor-deletion-vc}.

\begin{algorithm}[!ht]
\caption{{\problemname} parameterized by vertex cover}
\label{alg:f-minor-deletion-vc}
\begin{algorithmic}[1]
\Require A graph $G$ and a nonnegative integer $k$

\State Invoke \cite{DBLP:conf/stacs/0001N24} and compute the minimum vertex cover $X$. $I \gets V(G) \setminus X$.

\If{$k\ge |X|$}
    \State \Return \Yes
\EndIf

\ForAll{$D\subseteq X$ with $|D|\le k$}
    \State $Y\gets X\setminus D$ and $b\gets k-|D|$.

    \If{$G[Y]$ is not $\F$-minor-free}
        \State \textbf{continue}
    \EndIf

    \State $\Type(v)\gets N_G(v)\cap Y$ for every $v\in I$.
    \State $I_1\gets\{v\in I:|\Type(v)|\ge c_1+1\}$
        and $I_2\gets I\setminus I_1$.

    \If{$|I_1|>c_1|Y|+b$}\label{line:i1-branching}
        \State \textbf{continue}
    \EndIf

    \State $\Types\gets\{\Type(v):v\in I_2\}$.
    \State $B_Q\gets\{v\in I_2:\Type(v)=Q\}$
        for every $Q\in\Types$.

    \If{$|\Types|>c_2|Y|+b$}\label{line:types-branching}
        \State \textbf{continue}
    \EndIf

    \State $R_\emptyset\gets\emptyset$.
    \ForAll{$Q\in\Types$}\label{line:for-loop}
        \State Choose $R_Q\subseteq B_Q$
            according to \eqref{eq:vc-representatives}.
    \EndFor

    \State $R\gets\bigcup_{Q\in\Types}R_Q$.
    \State $G'\gets G[Y\cup I_1\cup R]$.

    \ForAll{$A\subseteq I_1$ and $Z\subseteq R$
        satisfying \eqref{eq:vc-admissible}}
        \If{$G'-(A\cup Z)$ is $\F$-minor-free}
            \State \Return \Yes
        \EndIf
    \EndFor
\EndFor

\State \Return \No
\end{algorithmic}
\end{algorithm}

\subsection{Correctness and Running Time}
\label{subsec:vc-correctness}

In this section, we prove the following theorem.

\begin{theorem}\label{thm:main-vc}
\Cref{alg:f-minor-deletion-vc} correctly decides {\problemname} and runs in $2^{O_\F(\vc(G))} \cdot n^{O(1)}$ time.
\end{theorem}

We first prove that if the algorithm returns {\Yes}, $D \cup A \cup Z$ is the solution in $G$.

\begin{lemma}\label{lem:vc-compression}
Fix a branch $D$ of \Cref{alg:f-minor-deletion-vc}, and let $(A,Z)$ be a pair enumerated in this branch.
Let $S:=D\cup A\cup Z$ and $C:=G'-(A\cup Z)$.
Then $C$ is $\F$-minor-free if and only if $G-S$ is $\F$-minor-free.
\end{lemma}

\begin{proof}
Since $C$ is a subgraph of $G-S$, if $G-S$ is $\F$-minor-free, then so is $C$.

For the converse, suppose that $C$ is $\F$-minor-free.
We show that all vertices in each $B_Q \setminus R_Q$ that are omitted in $C$ can be added back without creating a graph of $\F$ as a minor.

Fix a type $Q \in \Types$ such that $B_Q \setminus R_Q \neq \emptyset$.
If $Q \neq \emptyset$, then by the choice of $R_Q$ and the constraint on $Z$, $|R_Q \setminus Z| \ge |Q|(\Delta_\F + 1) + h - 1 - (|Q| - 1) = |Q| \Delta_\F + h$.
If $Q = \emptyset$, then $Z \cap R_\emptyset = \emptyset$ so that $|R_\emptyset| = h = |Q| \Delta_\F + h$.

All vertices of $Y$ remain, and $I$ is independent.
Thus, when the vertices of $B_Q\setminus R_Q$ are added back, the vertices of $B_Q\setminus Z$ are pairwise false twins with common neighborhood $Q$.
Since at least $|Q|\Delta_{\F}+h$ of them already belong to $C$, \Cref{lem:vc-false-twin-reduce} implies that adding the omitted vertices preserves $\F$-minor-freeness.

By applying this argument to every type $Q \in \Types$, we obtain $G - S$.
Therefore, $G - S$ is $\F$-minor-free.
\end{proof}

We next consider the solution $S^*$ chosen in \Cref{lem:vc-representatives}.
In its branch, each $R_Q$ contains enough vertices to choose $|S^*\cap B_Q|$ of them for deletion.
Since the vertices of $B_Q$ are false twins, this gives a graph isomorphic to $G-S^*$.

\begin{lemma}\label{lem:vc-solution-represented}
Suppose that $(G,k)$ is a \Yes-instance and $k<|X|$.
Let $S^*$ be chosen as in \Cref{lem:vc-representatives}.
Consider the branch $D:=S^*\cap X$.
Then there exist sets $A\subseteq I_1$ and $Z\subseteq R$ enumerated by \Cref{alg:f-minor-deletion-vc} such that $|A|+|Z|\le b$ and $G-(D\cup A\cup Z)$ is $\F$-minor-free.
\end{lemma}

\begin{proof}
The branch $D:=S^*\cap X$ is feasible.
Since $G[Y]$ is a subgraph of $G-S^*$, it is $\F$-minor-free.
By \Cref{lem:vc-high-deg-bound,lem:vc-type-bound}, the branch also satisfies both bounds on $|I_1|$ and $|\Types|$.
Hence, it is not discarded.

Let $A:=S^*\cap I_1$.
By \Cref{lem:vc-representatives}, $|S^* \cap B_\emptyset| = 0$ and for every nonempty $Q \in \Types$, $|S^* \cap B_Q| \le |Q| - 1$.
For every nonempty $Q\in\Types$, the definition of $R_Q$ gives
$|S^*\cap B_Q|\le \min\{|B_Q|,|Q|-1\}\le |R_Q|$.
Thus, we may choose $Z_Q\subseteq R_Q$ with $|Z_Q|=|S^*\cap B_Q|$.
Set $Z_\emptyset:=\emptyset$ and $Z:=\bigcup_{Q\in\Types}Z_Q$.

Then, $Z \cap R_\emptyset = \emptyset$ and $|Z \cap R_Q| \le |Q|-1$ for every nonempty $Q$.
Moreover,
\[
|A|+|Z|
=|S^*\cap I_1|+\sum_{Q\in\Types}|S^*\cap B_Q|
=|S^*\cap I|
\le k-|D|
=b.
\]
Thus $(A,Z)$ is enumerated by the algorithm.

For each $Q\in\Types$, the vertices of $B_Q$ are false twins in $G-D$, and $|Z_Q|=|S^*\cap B_Q|$.
Therefore, $G-(D\cup A\cup Z)\cong G-S^*$, so $G-(D\cup A\cup Z)$ is $\F$-minor-free.
\end{proof}

Now, we prove the main theorem of this section.

\begin{proof}[Proof of \Cref{thm:main-vc}]
Let $(G,k)$ be an input instance.
If $k\ge |X|$, deleting $X$ leaves an independent set.
Since every graph in $\F$ contains an edge, this is a solution,
and the algorithm correctly returns \Yes.
Hence, assume that $k<|X|$.

Suppose that the algorithm returns \Yes.
Then some enumerated pair $(A,Z)$ in a branch $D$ satisfies
$|D|+|A|+|Z|\le k$, and $G'-(A\cup Z)$ is $\F$-minor-free.
By \Cref{lem:vc-compression}, $G-(D\cup A\cup Z)$ is also
$\F$-minor-free.
Thus, $(G,k)$ is a \Yes-instance.

Conversely, suppose that $(G,k)$ is a \Yes-instance.
Choose $S^*$ as in \Cref{lem:vc-representatives}.
By \Cref{lem:vc-solution-represented}, the branch $D:=S^*\cap X$
enumerates a pair $(A,Z)$ for which $G-(D\cup A\cup Z)$
is $\F$-minor-free.
Its induced subgraph $G'-(A\cup Z)$ is also $\F$-minor-free,
so the algorithm returns \Yes.

It remains to analyze the running time.
Recall that $t=|X|=\vc(G)$.
Computing $X$ takes $2^{O(t)} \cdot |V(G)|^{O(1)}$ time~\cite{DBLP:conf/stacs/0001N24}.
If $k\ge t$, the algorithm returns immediately.
Otherwise, there are at most $2^t$ choices for $D$, and
each branch satisfies $|Y|\le t$ and $b=k-|D| \le k<t$.

Testing whether $G[Y]$ is $\F$-minor-free takes $2^{O_{\F}(t)}$ time by \Cref{lem:algo-minor-testing}.
Computing the sets $I_1$, $I_2$, and $B_Q$ and checking the two bounds takes polynomial time.

In a branch that is not discarded, we have $|I_1|\le c_1|Y|+b=O_{\F}(t)$ and $|\Types|\le c_2|Y|+b=O_{\F}(t)$.
Since $|Q|\le c_1$ for every $Q\in\Types$, each $R_Q$ contains at most $c_1(\Delta_{\F}+1)+h-1=O_{\F}(1)$ vertices.
Thus, $|R|=O_{\F}(t)$ and $|V(G')|=O_{\F}(t)$.

There are at most $2^{|I_1|+|R|}=2^{O_{\F}(t)}$ pairs $(A,Z)$ to consider.
For each pair, testing whether $G'-(A\cup Z)$ is $\F$-minor-free takes $2^{O_{\F}(t)}$ time by \Cref{lem:algo-minor-testing}.
Including the computation of $X$ and all choices of $D$, the total running time is $2^{O_{\F}(t)}\cdot |V(G)|^{O(1)} =2^{O_{\F}(\vc(G))}\cdot |V(G)|^{O(1)}$.
\end{proof}

Together with the case of families containing an edgeless graph, handled in \Cref{subsec:vc-setup}, we obtain the following theorem.

\mainupperbound*

\section{Conclusion}\label{sec:conclusion}

For every $h\ge 6$, we proved that, unless ETH fails,
\textsc{$\{P_h\}$-Minor-Free Deletion} cannot be solved in time
$2^{o(t\log t)}\cdot n^{O(1)}$,
where $t=\fvs(G)+\td(G)$.
It also yields lower bounds for each parameter separately.
Together with the algorithm of Baste, Sau, and
Thilikos~\cite{BasteST2023Hitting}, these bounds give tight
running-time dependence for all three parameterizations.
In contrast, we gave an algorithm solving
\textsc{$\mathcal{F}$-Minor-Free Deletion},
for every finite family $\mathcal{F}$, in time
$2^{O_{\mathcal{F}}(\vc)}\cdot n^{O(1)}$.
The latter running time cannot be improved in general, as already
for $\mathcal{F}=\{K_2\}$ the problem is \textsc{Vertex Cover},
which admits no $2^{o(\vc)}\cdot n^{O(1)}$-time algorithm
under ETH~\cite{DBLP:journals/jcss/ImpagliazzoPZ01}.
Thus, for paths on at least six vertices, the transition from
single-exponential to slightly super-exponential complexity happens
between the vertex cover number and the feedback vertex set number.
We conclude with several directions for further
research.

For a connected $H$, the complexity of \textsc{$\{H\}$-Minor-Free
Deletion} parameterized by treewidth is fully understood.
Baste, Sau, and Thilikos showed that, assuming the ETH,
the problem can be solved in time $2^{\Theta(\tw)}\cdot n^{O(1)}$ if $H$ is
a contraction of the \emph{chair} or the \emph{banner}, and requires time
$2^{\Theta(\tw\log\tw)}\cdot n^{O(1)}$
otherwise~\cite{BasteST2020Single,BasteST2020Hitting,BasteST2023Hitting}.

First, since $P_h$ for $h \le 4$ is a contraction of the chair, the problem is
single-exponential in $\tw$, and hence in $\fvs$,
for these
paths~\cite{BasteST2020Single,BasteST2023Hitting}.
Our results therefore leave
open only the case $h = 5$: can \textsc{$\{P_5\}$-Minor-Free Deletion} be
solved in time $2^{O(\fvs)}\cdot n^{O(1)}$, or does it also require time
$2^{\Omega(\fvs\log\fvs)}\cdot n^{O(1)}$ under the ETH?

More generally, it is natural to ask for an analogue of the above dichotomy
with respect to the feedback vertex set number.
Note that the above dichotomy does not extend to $\fvs$ (e.g., $H=K_4$~\cite{KimPP2015Single}).\footnote{$K_4$ is not a contraction of the chair or the banner, yet \textsc{$\{K_4\}$-Minor-Free Deletion} admits a single-exponential algorithm parameterized by~$\fvs$, since the single-exponential algorithm parameterized by
solution size is already known.}
Can we characterize all connected graphs $H$ for which
\textsc{$\{H\}$-Minor-Free Deletion} admits a single-exponential algorithm
parameterized by~$\fvs$?

Finally, the vertex-deletion distances to a linear forest
and to a star forest each lie between $\fvs$ and $\vc$.
For which families $\mathcal{F}$ does
\textsc{$\mathcal{F}$-Minor-Free Deletion} admit a
single-exponential algorithm parameterized by either of these distances?
In particular, does the lower bound for
$\mathcal{F}=\{P_h\}$ with $h\ge 6$ extend to either parameter?

\subparagraph*{Declaration of AI usage.}
The authors used Codex to edit language, prepare figures, and conduct literature searches. ChatGPT 5.5 Pro assisted with the initial development and subsequent generalization of the argument in \Cref{sec:fvs-lower-bound} from $h = 6$ to $h\geq 6$, as well as with identifying and correcting an error in \Cref{subsec:reduction}.
The authors independently verified all AI-assisted content and take full responsibility for the paper.
\bibliography{references}
\clearpage
\appendix
\section{Why the Construction Does Not Extend to $P_5$}
\label{app:path-five}

The restriction $h\ge 6$ is needed when checking the components
containing outputs with the same label.
In the proof of \Cref{prop:completeness}, several gadgets may
leave an output with label $(i,\pi(i))$.
These outputs are all adjacent to $r_i$ and $c_{\pi(i)}$.
If there are $s$ of them, they form a copy of $K_{2,s}$ together with two vertices $r_i$ and $c_{\pi(i)}$.

\begin{figure}[H]
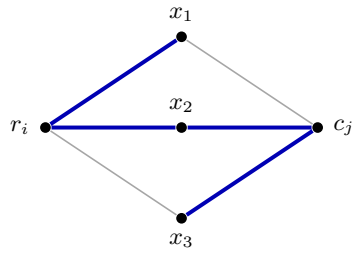

  \centering
  \RepeatedOutputsPfivePicture
  \caption{Three outputs with the same label $(i,j)$
    form a copy of $K_{2,3}$ with $r_i$ and $c_j$.
    The highlighted path $x_1-r_i-x_2-c_j-x_3$ is a $P_5$.}
  \label{fig:path-five-obstruction}
\end{figure}

For $h\ge 6$, we set $L=\lfloor(h-4)/2\rfloor\ge 1$.
Including the pendant paths of length $L$ at the two vertices,
the resulting component has no path on more than $2L+3<h$ vertices.
For $h=5$, the same choice gives $L=0$, so the pendant paths are absent.
However, if $s\ge 3$, the copy of $K_{2,s}$ already contains $P_5$.
Indeed, for three outputs $x_1,x_2,x_3$ with the same label $(i,j)$,
the sequence $x_1-r_i-x_2-c_j-x_3$ is a path on five vertices;
see \Cref{fig:path-five-obstruction}.

The problem cannot always be avoided by choosing different output vertices in the edge gadgets.
We give a yes-instance of {\permset} for which the construction with $h=5$ and $L=0$ produces a no-instance.

First, the tree construction in Section~\ref{sec:construction-base}
remains valid for $h=5$.
When $d=3$, each of the vertex-disjoint paths used in
\Cref{fig:packing-with-outputs,fig:packing-without-outputs}
has five vertices, and every component of $T_{5,q}-S_j$
is a path on at most four vertices.
Thus, the three properties of \Cref{lem:tree-properties}
also hold for the resulting trees $T_{5,q}$.

Now let $k=4$, and let $\Gamma$ have vertex set $[4]\times[4]$
and edge set
$E(\Gamma)=\{\{(i,j),(j,i)\}:1\le i<j\le 4\}$.
There are six edges.
The identity permutation gives the independent set
$\{(i,i):i\in[4]\}$, so $\Gamma$ is a yes-instance.

Apply the construction with $h=5$ and $L=0$,
and let $(G,B)$ be the resulting instance.
Suppose that $X$ is a $P_5$-deletion set of $G$ with $|X|\le B$.
The counting argument in \Cref{lem:deletions-per-gadget}
still applies: no vertex of $R\cup C$ is deleted,
and each of the four row gadgets and six edge gadgets
has an output vertex outside $X$.
Hence, at least ten output vertices remain.

If two remaining output vertices $x$ and $y$ have labels $(i,j)$ and $(i,j')$ with $j\ne j'$, then $c_j-x-r_i-y-c_{j'}$ is a path on five vertices in $G-X$, a contradiction.
Thus, for each row $i$, at most one label with first coordinate $i$ occurs among the remaining outputs.
Since there are four rows, at most four distinct labels occur.

Among at least ten remaining outputs with at most four labels, by the pigeonhole principle, three of them must have the same label.
They form a copy of $K_{2,3}$ with the corresponding vertices
of $R\cup C$, and hence give a $P_5$ in $G-X$.
This is a contradiction.

Thus, simply setting $h=5$ in the construction fails
even on a \Yes-instance, regardless of how the outputs are chosen.

\section{Computable Upper Bounds of Constants in \Cref{lem:prelim-linear-bound}}\label{app:constants}

For completeness, we present explicit computable upper bounds for the constants $c_1$ and $c_2$ used in \Cref{lem:prelim-linear-bound}.
We do not attempt to optimize these constants.

Fix a graph $H_0\in\F$ and let $t:=\max\{3,|V(H_0)|\}$.
Since $H_0$ is a subgraph of $K_t$, every $\F$-minor-free graph is $K_t$-minor-free.
By the theorem of Mader~\cite{Mader1968Homomorphiesatze}, every $K_t$-minor-free graph $G$ satisfies $|E(G)| \le 16t\log_2(t) \cdot|V(G)|$.
Hence, we may set $c_1:=16t\lceil\log_2(t)\rceil$.

For the constant $c_2$, we use the result of Gajarský, Hliněný, Obdržálek, Ordyniak, Reidl, Rossmanith, Sánchez Villaamil, and Sikdar~\cite{GajarskyHOORRSVS2017Kernelization}.
For a graph $G$, let $\nabla_1(G)$ denote the maximum edge density $|E(J)|/|V(J)|$ over all $1$-shallow minors $J$ of $G$.

\begin{lemma}[{\cite[Lemma~4.3]{GajarskyHOORRSVS2017Kernelization}}]\label{lem:app-gajarsky}
Let $G=(X,Y,E)$ be a bipartite graph with $X\ne\emptyset$,
and let $p$ be an integer greater than $\nabla_1(G)$.
Then, there are at most
\begin{enumerate}
    \item $2p \cdot |X|$ vertices of $Y$ with degree greater than $2p$;
    \item $(4^p + 2p) \cdot |X|$ distinct subsets $Q \subseteq X$ such that $Q = N_G(v)$ for some $v \in Y$.
\end{enumerate}
\end{lemma}

Since every $1$-shallow minor of an $\F$-minor-free graph is $\F$-minor-free, $\nabla_1(G)\le c_1$ for every $\F$-minor-free graph $G$.

Now let $G$ be an $\F$-minor-free graph and let $X\subseteq V(G)$ be nonempty.
Let $B$ be the bipartite subgraph of $G$ with bipartition $(X,V(G)\setminus X)$ containing precisely the edges crossing between $X$ and $V(G) \setminus X$.
Since $B$ is a subgraph of $G$, it is $\F$-minor-free, and hence $\nabla_1(B)\le c_1$.
Applying \Cref{lem:app-gajarsky} to $B$ with $p:=c_1+1$ gives
\[
    \left \lvert \{N_G(v)\cap X : v\in V(G)\setminus X\}\right \rvert
    \le \bigl(4^{c_1+1}+2(c_1+1)\bigr)|X|.
\]
Therefore, we may take $c_2:=4^{c_1+1}+2(c_1+1)$.

Both $c_1$ and $c_2$ are computable from $\F$.
Moreover, their binary representations have size polynomial in $t$.
Since $\F$ is fixed, their computation does not affect the running time of our algorithm.

\end{document}